\documentclass[journal]{IEEEtran}

\usepackage{graphicx}
\usepackage{cite}
\usepackage{picinpar}
\usepackage{amsmath}
\usepackage{url}
\usepackage{flushend}
\usepackage[latin1]{inputenc}
\usepackage{colortbl}
\usepackage{soul}
\usepackage{multirow}
\usepackage{pifont}
\usepackage{color}
\usepackage{alltt}
\usepackage[hidelinks]{hyperref}
\usepackage{enumerate}
\usepackage{siunitx}
\usepackage{breakurl}
\usepackage{epstopdf}
\usepackage{pbox}
\usepackage{amsfonts}
\usepackage{amssymb}
\usepackage{booktabs}
\usepackage{mathrsfs}
\usepackage[caption=false,font=normalsize,labelfont=sf,textfont=sf]{subfig}

\usepackage{tikz}

\newcommand\copyrighttext{%
  \footnotesize This work has been submitted to the IEEE for possible publication. Copyright may be transferred without notice, after which this version may no longer be accessible.}
\newcommand\copyrightnotice{%
\begin{tikzpicture}[remember picture,overlay]
\node[anchor=south,yshift=10pt] at (current page.south) {\fbox{\parbox{\dimexpr\textwidth-\fboxsep-\fboxrule\relax}{\copyrighttext}}};
\end{tikzpicture}%
}

\newtheorem{definition}{Definition}
\newtheorem{theorem}{Theorem}
\newtheorem{lemma}{Lemma}
\newtheorem{problem}{Problem}

\newtheorem{corollary}{Corollary}

\begin{document}

	\title{Privacy-Preserving Fusion for Multi-Sensor Systems Under Multiple Packet Dropouts}

	\author{
		Jie~Huang,~Jason~J.~R.~Liu,~\IEEEmembership{Member,~IEEE}~and~Xiao~He,~\IEEEmembership{Senior~Member,~IEEE} 
		\thanks{This work is supported in part by the National Natural Science Foundation of China under Grant 62403008 and Grant 62473223, the Macao Science and Technology Development Fund under Grant 0139/2023/RIA2, the University of Macau under Grant MYRG-CRG2024-00037-FST-ICI, Guangdong Basic and Applied Basic Research Fund under Grant 2025A1515012309, and Beijing Natural Science Foundation under Grant L241016. (Corresponding author: Jason J. R. Liu.)}
		\thanks{Jie Huang and Jason J. R. Liu are with the Department of Electromechanical Engineering, and the Centre for Artificial Intelligence and Robotics, University of Macau, Macau, China. Xiao He is with the Department of Automation, Tsinghua University, Beijing 100084, P.~R.~China. 
			(e-mail: huangjie18@mails.tsinghua.edu.cn;~jasonliu@um.edu.mo;~hexiao@tsinghua.edu.cn)}
		
	}

	\markboth{}
	{}

	\maketitle
        \copyrightnotice
	
	\begin{abstract}
	Wireless sensor networks (WSNs) are critical components in modern cyber-physical systems, enabling efficient data collection and fusion through spatially distributed sensors. However, the inherent risks of eavesdropping and packet dropouts in such networks pose significant challenges to secure state estimation. In this paper, we address the privacy-preserving fusion estimation (PPFE) problem for multi-sensor systems under multiple packet dropouts and eavesdropping attacks. To mitigate these issues, we propose a distributed encoding-based privacy-preserving mechanism (PPM) within a control-theoretic framework, ensuring data privacy during transmission while maintaining the performance of legitimate state estimation. A centralized fusion filter is developed, accounting for the coupling effects of packet dropouts and the encoding-based PPM. Boundedness conditions for the legitimate user's estimation error covariance are derived via a modified algebraic Riccati equation. Additionally, by demonstrating the divergence of the eavesdropper's mean estimation error, the proposed PPFE algorithm's data confidentiality is rigorously analyzed. Simulation results for an Internet-based three-tank system validate the effectiveness of the proposed approach, highlighting its potential to enhance privacy without compromising estimation accuracy.
		
	\end{abstract}


	\IEEEpeerreviewmaketitle

	\section{Introduction}
	
	WSNs collect information using smart sensors that are spatially distributed and wirelessly connected. These sensors transmit data to a fusion center, where it is processed to generate supervision over systems, thereby reducing wiring complexity and enhancing scalability. Consequently, fusion estimation for multi-sensor systems has been extensively studied across various industries, including cyber-physical systems \cite{7428772}, the Internet of Things \cite{CENGIZ2025103102}, target tracking systems \cite{8957406,7467492}, and environmental monitoring systems \cite{8283521}.
	In a typical sensor network, the broadcast nature of wireless communication inevitably attracts malicious attackers \cite{ZHANG2024111408,10242077,11028075}. Eavesdropping is one such cyber threat, potentially compromising the privacy of sensitive data such as measurements, states, and control inputs \cite{Kogiso2015_C,10440469,9991185,10750420}. This has led to significant research interest in the PPFE problem to counter eavesdropping, resulting in numerous important contributions \cite{10552563,9940579,9612083}.
	
	Due to the computational and communication resource limitations in WSNs, cryptography-based privacy-preserving frameworks, while offering the highest level of secrecy, are often impractical. To address these constraints, control-theoretic secrecy approaches have emerged as an efficient alternative \cite{Tsiamis2017_P,Leong2017_O}. These methods utilize the expected error covariance as a performance metric to ensure confidentiality. By adopting this framework, PPMs can be seamlessly integrated into control tasks, taking into account the physical layer of wireless communication.
	In this context, encoding-based PPM methods have been developed to safeguard data privacy during wireless transmission, offering a low computational burden and relatively high data availability. In \cite{Tsiamis2017_P}, an encoding-based method for unstable systems have been designed to achieve secrecy by stochastically withholding data. For stable systems, an encoding scheme utilizing pseudo-random noise transmission has been proposed to counter eavesdroppers in \cite{CRIMSON20238363}. Additionally, \cite{Zou2023_E} demonstrates the confidentiality performance of estimation error at the eavesdropper's side and introduces an auxiliary system-based encoding-decoding scheme.
	
	Packet dropout, a common phenomenon in communication networks, frequently occurs in networked control systems, complicating the balance between data availability and confidentiality in PPM design. Some efforts have been made to address the degradation of estimation performance on the legitimate user's side. For instance, a perfect secrecy encoding scheme for both stable and unstable systems with packet dropouts is proposed in \cite{Tsiamis2017_S, Tsiamis2018_S}, introducing the concept of critical events for lossless encoding. To mitigate the high computational burden caused by matrix power operations in perfect secrecy encoding, an expectational lossless PPM was introduced in \cite{Huang2022_P}.
	The aforementioned works focus on single packet dropout rates. However, in multi-sensor systems, data packets from different sensors may experience varying dropout probabilities during transmission. Despite this, very few studies have addressed the PPFE problem under the control-theoretic framework in the presence of multiple packet dropouts. This lack of research is primarily due to the mathematical challenges associated with analyzing the coupling between multiple packet dropouts and encoding-based PPMs. Addressing these challenges serves as the primary motivation for our current research.
	
	In this paper, we endeavor to develop a new PPFE algorithm for a general class of multi-sensor systems that incorporates the challenges posed by multiple packet dropouts and eavesdropping attacks. The main contributions of this work are summarized as follows:
	(1) For the first time, we formulate and solve an encoding-based PPFE problem within a control-theoretic framework, facilitating the mitigation of the combined adverse effects of multiple packet dropouts and eavesdropping attacks.
	(2) To ensure data availability, we derive upper bounds and establish boundedness conditions on the legitimate user's estimation error covariance by constructing a modified algebraic Riccati equation that fully accounts for the coupling effects between the PPM and multiple packet dropouts.
	(3) The secrecy performance is adequately analyzed by demonstrating that the eavesdropper's expected estimation error diverges under the proposed PPFE design.

	The remainder of this paper is organized as follows. Section II formulates the PPFE design problem under the scenario of multiple packet dropouts. In Section III, a centralized privacy-preserving fusion filter is developed, and its secrecy is validated through performance analysis of both the legitimate user's and the eavesdropper's estimation errors. Section IV presents a numerical example demonstrating the effectiveness of the proposed filtering framework, while Section V lastly draws a few conclusions.

	\textit{Notations:} 
	$\mathbb{R}^{d_x}$ is the $d_x$-dimensional Euclidean space. 
	$Y \le X$ denotes that the matrix $\left( X-Y \right) $ is positive semidefinite.
	For $X \in \mathbb{R}^{m\times n}$, $X^{\top}$ is the transpose of $X$.
	The maximum and minimum eigenvalues of a matrix are denoted by $\lambda _{\max} \left\{ \cdot  \right\} $ and $\lambda _{\min} \left\{ \cdot  \right\}$, respectively.
	$\mathrm{diag} \left\{ y \left( i \right) \right\}_{i=1} ^{M}$ denotes the diagonal matrix with the diagonal elements $ y \left( 1\right), y \left( 2 \right), \dots, y \left( M\right) $; 
	$I_{n}$ is the $n$-dimensional identity matrix.
	$\left\{ \gamma_{k,m} \right\}_{\mathbb{N}_0}$ is the sequence $\left\{\gamma_0, \gamma_1, \cdots \right\}$.
	$\left\| Y \right\| $ denotes the Euclidean norm of a vector or a matrix $Y$.
	$\mathcal{N} \left(0, \Sigma \right) $ denotes the normal distribution with zero-mean and covariance $\Sigma$.
	$\mathrm{Pr} \left\{ \mathcal{E} \right\} $  is the probability of the event $\mathcal{E}$.
	$\mathbb{E} \left[ X \right] $ and $\mathrm{Cov} \left(  X \right) $ denote the expectation and variance of a random variable $X$.
	
	\section{Problem Statement}
	
\begin{figure}[!htbp]
		\centering
		\includegraphics[width=1\linewidth,scale=1]{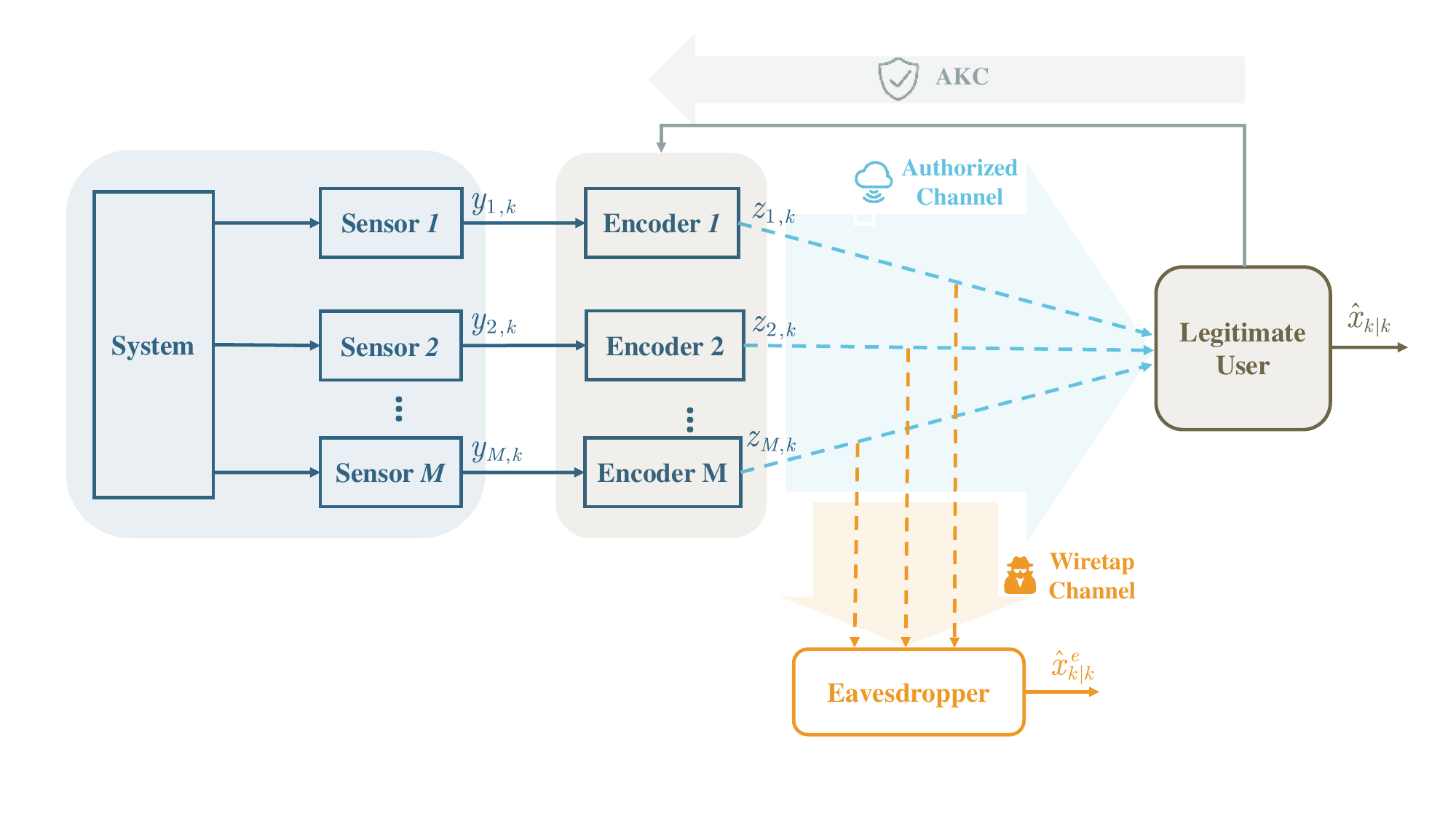}
		\caption{The fusion estimation architecture in presence of an eavesdropper}
		\label{Chap04:Fig1}
	\end{figure}

	Consider the design of a centralized PPFE for sensor networks, as illustrated in Fig. \ref{Chap04:Fig1}. Within this framework, multiple packet dropouts may occur during data transmission in the communication channels. Simultaneously, an eavesdropper may attempt to intercept the transmitted data. 
	In what follows, we will detailedly introduce the system plant, the communication channel model, and the PPM structure.
	
	\subsection{Multi-Sensor Systems With Multiple-Packet Transmissions}
	
	Consider a linear discrete-time system, with states observed by $M$ sensors, can be described as follows:
	\begin{align}\label{chap04:eq1}
		& \begin{cases}	x_{k+1}=Ax_k+w_{k+1},\\	y_{i,k}\,\,  =C_ix_k+v_{i,k},\\\end{cases} \nonumber \\
		& ~~~ i=1, 2, \cdots, M,
	\end{align}
	where  $x_k \in \mathbb{R} ^{d_x}$ is the state vector, with the initial state $x_0\sim\mathcal{N} \left( \bar{x}_0,\bar{P}_0 \right) $, $\bar{P}_0\in \mathbb{R} ^{d_x \times d_x} \ge 0$;
	$y_{i,k} \in \mathbb{R} ^{d_{y_i}}$ represents the measurement data collected by the $i$-th sensor. 
	$w_k\in \mathbb{R} ^{d_x}$ and $v_{i,k} \in \mathbb{R} ^{d_{y_i}}$ are zero-mean white Gaussian noises and satisfy the following statistical relationships:
	(1) $ \mathbb{E} [ w_k w_l^\top ] = \mathbf{1}_{( k = l)} Q $, with $Q \in \mathbb{R} ^{d_x \times d_x} \ge 0$;
	(2) $ \mathbb{E} [ v_{i,k} v_{i,l}^\top ] = \mathbf{1}_{( k = l)} R_i $, with $R_i\in \mathbb{R} ^{d_{y_i} \times d_{y_i}} > 0$;
	(3) $ \mathbb{E} [ w_k v_{i,l}^\top ] = 0 $, which indicates $w_k$ and $v_{i,l} $ arethat  mutually independent;
	(4) $ \mathbb{E} [ v_{i,k} v_{j,k}^\top ] = \mathbf{1}_{( i = j)} R_i $;
	where
	\begin{equation}
		\mathbf{1}_{(k = l)} =
		\begin{cases}	1, ~~~ \textit{if} ~k=l, \\	0,~~~\textit{otherwise} .\\ \end{cases} \nonumber
	\end{equation}
	Assume the parameter matrix $C_i$ has full row rank. The  matrices $A$, $C_i$, $Q$, $R_i$, and $\bar{P}_0$	are publicly known to all entries.
	
	The measurement data is transmitted through $M$ communication channels to a remote legitimate user for state estimation. 
	Due to limited bandwidth of the communication channels, the data packet transmitted through different channels may be dropped with different probabilites.
	Also, the broadcasting nature of communication network leads to the potential risks of eavesdropping.
	
	The channel output can be described as follows:
	\begin{equation}\label{chap04:eq3}
		\zeta _{i,k}=\begin{cases}	z_{i,k}, ~\text{if}~ \,\,\gamma _{i,k}=1,\\	0,  ~~~~\text{if}~ \,\,\gamma _{i,k}=0,\\\end{cases}\,\,\zeta _{i,k}^{e}=\begin{cases}	z_{i,k},  ~\text{if}~ \,\,\gamma _{i,k}^{e}=1,\\	0,  
			~~~~\text{if}~ \,\,\gamma _{i,k}^{e}=0,\\\end{cases}\,\,
	\end{equation}
	where $\zeta _{i,k},~ \zeta _{i,k}^{e} \in \mathbb{R} ^{d_{y_i}}$ respectively represent the received measurements at the legitimate user's and the eavesdropper's side via the $i$-th channel. 
	$z_{i,k} \in \mathbb{R} ^{d_{y_i}}$ is the encoded measurements against eavesdropper, where the encoding mechanism will be introduced in the next subsection.
	The Bernoulli processes $ \{ \gamma _{i,k} \} $ and $ \{ \gamma _{i,k}^{e} \} $ represent the outcomes of the authorized and wiretap channels, separately.
	The probabilities for data packets to be successfully received or wiretapped can be defined as:
	\begin{align}\label{chap04:eq4}
		& \bar{\gamma}_i=\mathrm{Pr}\left\{ \gamma _{i,k}=1 \right\} = \mathbb{E} \left[ \gamma _{i,k} \right] , \nonumber \\ 
		& \bar{\gamma}_i^{e}=\mathrm{Pr}\left\{ \gamma _{i,k}^{e}=1  \right\} = \mathbb{E} \left[ \gamma _{i,k}^{e} \right],
	\end{align}
	where $\bar{\gamma}_i,\bar{\gamma}_i^{e}\in \left( 0,1 \right] $, and $\lim_{k\rightarrow \infty} \bar{\gamma}_i\ne 0$, $\lim_{k\rightarrow \infty} \bar{\gamma}_i^{e}\ne 0$.
	To depict the effect of the multiple packet droppout, the concept of channel capacity is introduced.
	
	The individual channel capacity of the
	$i$-th channel is defined as:
	\begin{equation}\label{chap04:eq5}
		\mathscr{C} _i\triangleq \frac{1}{2}\ln \left( 1+\frac{\bar{\gamma}_i}{1-\bar{\gamma}_i} \right) =-\frac{1}{2}\ln \left( 1-\bar{\gamma}_i \right).
	\end{equation}
	Correspondingly, the overall channel capacity is defined as:
	\begin{equation}\label{chap04:eq6}
		\mathscr{C} _k \triangleq \sum_{i=1}^M{\mathscr{C} _i}=-\sum_{i=1}^M{\frac{1}{2}\ln \left( 1-\bar{\gamma}_i \right)}.
	\end{equation}

	\subsection{Encoding-Decoding-Based PPM}
	For the PPM, a decentralized architecture is adopted, where local PPM designs are improved based on the approach established in the \cite{Huang2022_P}. 
	The PPM encodes the measurements using the following encoding function:
	\begin{equation}\label{chap04:eq7}
		z_{i,k}=\mathcal{Q} _i\left( \frac{y_{i,k}-\left( a_i \right) ^{k-t_{i,k}}\bar{y}_{t_{i,k}}}{s} \right), 
	\end{equation}
	where $a_i > 1$ and $s \neq 0$ are scalar parameters of the encoding mechanism. 
	The parameter $a_i$ can accelerate the divergence rate of the eavesdropper's estimation error, while  $s$
	can regulate the length of the transmitted data and adjust the covariance of the decoding error. 
	Since the adjustment brought by  $s$ affects both the legitimate user and the eavesdropper's estimation results, the design of this parameter requires a trade-off. 
	The reference time $t_{i,k}$ represents the most recent time when the channel successfully transmits the data, which can be defined 
	based on the outcomes of each authorized channel:
	$$t_{i,k}=\max \left\{ t:0\le t<k, \gamma _{i,k}=1 \right\} .$$	
	Let us define the input of the mapping function $\mathcal{Q} _i\left( \cdot \right) $ as $\bar{z}_{i,k}= \left[ y_{i,k}-\left( a_i \right) ^{k-t_{i,k}}\bar{y}_{t_{i,k}} \right] / s$.
	When the entry $\bar{z}_{i,k} \left( \ell \right) \in \left[ d_i\delta_i, \left( d_i +1 \right) \delta_i \right] $, for $ \ell=1,2,\cdots,d_{y_i}$, the measurement data $y_{i,k} \left( \ell \right) $ is encoded using the a probabilistic uniform quantization function $\mathcal{Q} _i\left( \cdot \right) $, following the rule:
	\begin{equation}\label{chap04:eq8}
		\begin{cases}	\mathrm{Pr} \left\{ z_{i,k}\left( \ell \right) =d_i\delta _i \right\} =1-q_i,\\	\mathrm{Pr} \left\{ z_{i,k}\left( \ell \right) =\left( d_i+1 \right) \delta _i \right\} =q_i,\\\end{cases}
	\end{equation}
	where the quantization probability is defined as:
	$$q_i=\left[ \bar{z}_{i,k} \left( \ell \right)  -d_i\delta _i \right] /\delta _i\in \left[ 0,1 \right] .$$	
	Thus, the encoded measurement data is mapped to the set $\mathcal{E} _i=\left\{ d_i\delta _i, d_i\in \mathbb{Z} , \delta _i>0 \right\} $.
	The encoding error is defined as $e_{i,k}=z_{i,k}- \bar{z}_{i,k} $, satisfying:
	$$\begin{cases}	\mathrm{Pr} \left\{ e_{i,k} \left( \ell \right) =-q_i\delta _i \right\} =1-q_i,\\	\mathrm{Pr} \left\{ e_{i,k}\left( \ell \right) =\left( 1-q_i \right) \delta _i \right\} =q_i.\\\end{cases}$$	
	The quantization error has the following statistical properties: $\mathbb{E} \left[ e_{i,k} \left( \ell \right) \right] =0$, 
	$\mathrm{Var} \left( e_{i,k} \left( \ell \right) \right) =q_i\left( 1-q_i \right) \delta _{i}^{2}\le \frac{\delta_{i}^2}{4} $.
	Since the quantization are independently carried out for each local component, the error $\mathbb{E} \left[ e_{i,k} e_{j,k} ^\top \right] =0$. 
	Also, the quantization error and the measurement noise satify that $\mathbb{E} \left[ e_{i,k} v_{j,k} ^\top \right] =0$.
	For the encoding mechanism \eqref{chap04:eq7}-\eqref{chap04:eq9}, the decoding mechanism is defined as:
	\begin{equation}\label{chap04:eq9}
		\bar{y}_{i,k}=z_{i,k}s+\left( a_i \right) ^{k-t_{i,k}}\bar{y}_{t_{i,k}}.
	\end{equation}
	Let decoding error define as $e_{i,k} ^{dec}  =\bar{y}_{i,k}- y_{i,k} $. We can calculate that 
	$\mathbb{E} \left[ e_{i,k} ^{dec} \left( \ell \right) \right] = s \mathbb{E} \left[ e_{i,k} \left( \ell \right) \right] = 0$ and 
	$\mathrm{Var} \left( e_{i,k} ^{dec} \left( \ell \right) \right) = s^2 \mathrm{Var} \left( e_{i,k} \left( \ell \right) \right) = s^2 q_i\left( 1-q_i \right) \delta _{i}^{2}\le  \frac{ s^2 \delta_{i}^2}{4} $
	Therefore, this PPM is lossless in expectation.

	\subsection{Problem}
	
	The goal of this paper is to design an information fusion filtering algorithm that satisfies the secrecy requirements as defined below.
	
	\begin{definition}\label{definition1}
		Given a multi-sensor system \eqref{chap04:eq1} with communication channels \eqref{chap04:eq3}-\eqref{chap04:eq6}, the fusion filtering algorithm under the PPM described in \eqref{chap04:eq7}-\eqref{chap04:eq9} achieves secrecy if and only if both the following two conditions are satisfied:
		
		\begin{itemize}
			\item[(i)] The covariance matrix of the legitimate user's estimation error, $\tilde{x}_{k|k}$, is bounded, 
			i.e. there exists a matrix $\bar{P}^{\star} >0 $ such that
			$\mathbb{E} \left[ \tilde{x}_{k|k} \tilde{x}_{k|k}^\top  \right] \leq \bar{P}^{\star}$.
			\item[(ii)] The expected value of the eavesdropper's estimation error, $\tilde{x}_{k|k}^{e}$, is divergent,
			i.e. 
			$\lim_{k\rightarrow \infty} \left\| \mathbb{E} \left[ \tilde{x}_{k|k}^{e} \right] \right\| =\infty$.
		\end{itemize}
	\end{definition}

This definition establishes the conditions for data availability and confidentiality. Specifically, it ensures that the legitimate user has bounded estimation error covariance, while the expected value of the eavesdropper's estimation error diverges. This implies that private data, such as the system state, cannot be reconstructed by an eavesdropper but remains accessible and usable by the legitimate user.

\begin{problem}
	Consider the multi-sensor system \eqref{chap04:eq1}, where information leakage occurs over the communication channels \eqref{chap04:eq3}-\eqref{chap04:eq6}. Under the PPM \eqref{chap04:eq7}-\eqref{chap04:eq9}, design a privacy-preserving fusion estimator that satisfies the secrecy requirements outlined in Definition \ref{definition1}.
\end{problem}

	\section{Main Results}
	
	In this section, we will first develop a centralized fusion estimator under the PPM \eqref{chap04:eq7}-\eqref{chap04:eq9} while accounting for multiple packet dropouts. Subsequently, we will evaluate the secrecy of the proposed PPFE algorithm by assessing the estimation performance of both the legitimate user and the eavesdropper. Specifically, we determine the boundedness conditions for the legitimate user's actual estimation error covariance and verify the divergence of the eavesdropper's mean estimation error.
	
	\subsection{Centralized Privacy-Preserving Fusion Estimator}
	Within the privacy definition framework based on state estimation error, we will present a privacy-preserving fusion estimation under multiple-packet-dropout channels.
	
	Let us define the argumented vectors and matrices as follows:
	\begin{align}\label{chap04:eq10}
		& \breve{y}_k \triangleq \left[ \begin{matrix} \gamma _{1,k} \bar{y}_{1,k} ^\top & \gamma _{2,k} \bar{y}_{2,k} ^\top &  \cdots & \gamma _{M,k} \bar{y}_{M,k} ^\top \\\end{matrix} \right]  ^\top , \nonumber   \\
		& \breve{C}_k \triangleq \left[ \begin{matrix} \gamma _{1,k} C_1 ^\top  &  \gamma _{2,k} C_2 ^\top  &  \cdots & \gamma _{M,k} C_M ^\top \\\end{matrix} \right]  ^\top,  \nonumber   \\
		& \breve{v}_k \triangleq \left[ \begin{matrix} \gamma _{1,k} v_{1,k} ^\top & \gamma _{2,k} v_{2,k} ^\top &  \cdots  &  \gamma _{M,k} v_{M,k} ^\top \\\end{matrix} \right] ^\top, \nonumber   \\
		& \breve{e}_k \triangleq \left[ \begin{matrix} \gamma _{1,k} e_{1,k} ^\top & \gamma _{2,k} e_{2,k} ^\top &  \cdots &  \gamma _{M,k} e_{M,k} ^\top \\\end{matrix} \right] ^\top, \nonumber   \\
		& \breve{z}_k \triangleq \left[ \begin{matrix} \gamma _{1,k} z_{1,k} ^\top & \gamma _{2,k} z_{2,k} ^\top & \cdots &  \gamma _{M,k} z_{M,k} ^\top \\\end{matrix} \right] ^\top,  \nonumber   \\
		& \vec{y}_k \triangleq \left[ \begin{matrix} \gamma _{1,k} y_{1,k} ^\top & \gamma _{2,k} y_{2,k} ^\top &  \cdots &  \gamma _{M,k} y_{M,k} ^\top \\\end{matrix} \right] ^\top, \nonumber \\
		& \breve{e}_{dec,k} \triangleq s \left[ \begin{matrix} \gamma _{1,k} e_{1,k}^\top & \gamma _{2,k} e_{2,k}^\top &  \cdots &  \gamma _{M,k} e_{M,k}^\top \\\end{matrix} \right] ^\top. 
	\end{align}
	The measurement equation for the remote estimator is derived as:
	\begin{equation}\label{chap04:eq11}
		\breve{y}_k=\breve{C}_kx_k+\breve{v}_k+\breve{e}_{dec,k}.
	\end{equation}
	
	Due to the nonlinearity introduced by the probabilistic mapping function $\mathcal{Q} _i\left( \cdot \right)$, the decoded measurement $ \bar{y}_{i,k}$ may not follow a Gaussian distribution. 
	However, the properties of probabilistic uniform quantization, as indicated in \cite{Ribeiro_TSP2006_S,You_ACC2009_R}, suggest that the distribution of the decoded measurement will closely approximate a Gaussian distribution as the number of encoding levels increases. 
	Given that the initial state $ x_0 $ follows a Gaussian distribution, we assume that the one-step predicted state distribution also approximates a Gaussian distribution under the PPM in Eqs. \eqref{chap04:eq7}-\eqref{chap04:eq9}. 
	During the design of the fusion filter, the correlation between the encoding error $e_{i,k}$ and the measurement noise $v_{i,k}$ is considered negligible. Consequently, we can derive a centralized fusion filter with a structure similar to that of a Kalman filter, which achieves a recursive structure, thereby increasing feasibility and reducing computational complexity. 
	However, the encoding error may degrade fusion performance. Therefore, we will analyze the actual estimation error covariance in the next section, where the correlation between the encoding error and the measurement noise will be considered.
	
	We define a new vector $\mathfrak{v} _{i,k}=\left[ \begin{matrix}	v_{i,k}^\top &		e_{i,k}^\top \\\end{matrix} \right] ^\top $ and assume that
	$\mathfrak{v} _{i ,k}$ and $\mathfrak{v} _{j ,k}$ are uncorrelated when $ i\ne j $.
	Thus, the argumented measurement noise and decoding error covariance matrices are $ \breve{R}=\mathrm{diag}\left\{ \left( \gamma _{i,k} \right) ^2R_i \right\} _{i=1}^{M} $ and $ \breve{R}_{dec} =\mathrm{diag}\left\{ \left( \gamma _{i,k} \right) ^2s^2 R_{e_{i,k}} \right\} _{i=1}^{M} $, respectively. The local quantization error covariance $R_{e_{i,k}}  = \mathrm{diag}\left\{ \mathrm{Var} \left( e_{i,k} \left( \ell \right) \right) \right\} _{i=1}^{M}$.
	
	\begin{theorem}\label{chap04:theorem1}
		For the multi-sensor system \eqref{chap04:eq1} with data encoded via  \eqref{chap04:eq7}-\eqref{chap04:eq9} and transmitted via  \eqref{chap04:eq3}-\eqref{chap04:eq6}, the predicton and measurement update equations of the remote fusion estimator are shown as follows:
		\begin{align}
			\hat{x}_{k|k-1}&=A\hat{x}_{k-1|k-1}, \label{chap04:eq13}\\ 
			\hat{x}_{k|k}& =\hat{x}_{k|k-1}+K_k\left( \breve{y}_k-\breve{C}_k\hat{x}_{k|k-1} \right),  \label{chap04:eq14}
		\end{align}
		where the filter gain is $K_k=P_{k|k-1}\breve{C}_k^\top \left( \breve{C}_kP_{k|k-1}\breve{C}_k^\top +\breve{R} \right) ^{-1}$. The corresponding error covariance matrices are:
		\begin{align}
			& P_{k|k-1} =AP_{k-1|k-1}A^\top +Q, \label{chap04:eq15} \\
			& P_{k|k} =P_{k|k-1}-K_k\left( \breve{C}_kP_{k|k-1}\breve{C}_k^\top +\breve{R} \right) K_{k}^\top \nonumber \\
			& ~~~~~~~~~+K_k\left( s^2 \breve{R}_{dec} \right) K_{k}^\top .  \label{chap04:eq16}
		\end{align}
	\end{theorem}
	\begin{IEEEproof}
		Since the $\sigma $-algebra generated by $\left\{ \zeta _{1:k-1},\gamma _k,z_k \right\} $ is a sub-$\sigma $-algebra of that generated by 
		$\left\{ \zeta _{1:k-1},\gamma _k,\varepsilon _k \right\} $, the state estimation can be derived as:
		\begin{align}\label{chap04:eq17}
			\hat{x}_{k|k}=\mathbb{E} \biggl\{ \mathbb{E} \left[ x_k|\left\{ \gamma _{i,k} \right\} _{i=1}^{M},\left\{ z_{i,k} \right\} _{i=1}^{M},\left\{ y_{i,1:k-1} \right\} _{i=1}^{M} \right]  \nonumber \\
			\bigg| \left\{ \gamma _{i,k} \right\} _{i=1}^{M},	\left\{ z_{i,k} \right\} _{i=1}^{M},\left\{ z_{i,1:k-1} \right\} _{i=1}^{M} \biggr\} . 
		\end{align}
		The state estimation for the centralized Kalman filter is then defined as
		\begin{align}\label{chap04:eq18}
			\hat{x}_{k|k}^{KF} & \triangleq \mathbb{E} \left[ x_k \bigg| \left\{ \gamma _{i,k} \right\} _{i=1}^{M},\left\{ z_{i,k} \right\} _{i=1}^{M},\left\{ y_{i,1:k-1} \right\} _{i=1}^{M} \right] \nonumber \\
			&  =\hat{x}_{k|k-1}+K_k\left( \vec{y}_k-\breve{C}_k\hat{x}_{k|k-1} \right) ,
		\end{align}
		where the gain matrix $K_k=P_{k|k-1}\breve{C}_k^\top \left( \breve{C}_kP_{k|k-1}\breve{C}_k^\top +\breve{R} \right) ^{-1}$.
		Consequently, we obtain that
		\begin{equation}\label{chap04:eq19}
			\hat{x}_{k|k}=\hat{x}_{k|k-1}+K_k\left( \breve{y}_k-\breve{C}_k\hat{x}_{k|k-1} \right) .
		\end{equation}
		The corresponding state estimation error covariance is derived as follows:
		\begin{align}\label{chap04:eq20}
			P_{k|k} & =\mathbb{E} \biggl\{ \mathbb{E} \bigg[ \left( x_k-\hat{x}_{k|k} \right) \left( x_k-\hat{x}_{k|k} \right) ^\top \bigg| \left\{ \gamma _{i,k} \right\} _{i=1}^{M}, \nonumber  \\
			&~~~~~~~~~~~~~~~~~~~~~~~~~~~~~~~~~~ \left\{ z_{i,k} \right\} _{i=1}^{M},\left\{ y_{i, 1:k-1} \right\} _{i=1}^{M} \bigg] \biggr\} \nonumber \\
			& =\mathbb{E} \bigg[ \left( x_k-\hat{x}_{k|k}^{KF} \right) \left( x_k-\hat{x}_{k|k}^{KF} \right) ^\top \bigg| \left\{ \gamma _{i,k} \right\} _{i=1}^{M},  \nonumber  \\
			&~~~~~~~~~~~~~~~~~~~~~~~~~~~~~~~~~~  \left\{ z_{i,k} \right\} _{i=1}^{M},\left\{ y_{i, 1:k-1} \right\} _{i=1}^{M} \bigg]  \nonumber  \\
			& ~~~~ +\mathbb{E} \bigg[ \left( \hat{x}_{k|k}^{KF}-\hat{x}_{k|k} \right) \left( \hat{x}_{k|k}^{KF}-\hat{x}_{k|k} \right) ^\top  \nonumber  \\
			&~~~~~~~~~~~~~~~~~~~~~~\bigg| \left\{ \gamma _{i,k} \right\} _{i=1}^{M} ,  \left\{ z_{i,k} \right\} _{i=1}^{M},\left\{ y_{i, 1:k-1} \right\} _{i=1}^{M} \bigg]  \nonumber  \\
			& =P_{k|k-1}-K_k\left( \breve{C}_kP_{k|k-1}\breve{C}_k^\top +\breve{R} \right) K_{k}^\top   \nonumber  \\
			& ~~~~+\mathbb{E} \left[ \bar{K}_k\left( \vec{y}_k-\breve{y}_k \right) \left( \vec{y}_k-\breve{y}_k \right) ^\top \bar{K}_{k}^\top  \right]  \nonumber  \\
			& =P_{k|k-1}-K_k\left( \breve{C}_kP_{k|k-1}\breve{C}_k^\top +\breve{R} \right) K_{k}^\top  \nonumber  \\
			&~~~~ +K_k \left( s^2 \breve{R}_{dec} \right) K_{k}^\top .
		\end{align}
	\end{IEEEproof}

	\subsection{Estimation Performance for the Legitimate User}
	
	This subsection analyzes the estimation performance of legitimate users under the PPM scheme described in (\ref{chap04:eq7}-\ref{chap04:eq8}). Unlike the previous subsection, where the quantization error $e_{i,k}$ and measurement noise $v_{i,k}$ were assumed uncorrelated in the derivation of the estimation error covariance $P_{k|k}$, here we consider the actual estimation error covariance, thereby accounting for their correlation.
	The estimation errors are defined as $ \tilde{x}_{k|k-1} \triangleq x_k-\hat{x}_{k|k-1} $ and $ \tilde{x}_{k|k} \triangleq x_k-\hat{x}_{k|k} $,
	which leads to the actual error covariance matrices:
	\begin{align}\label{chap04:eq22}
		\nonumber \Sigma _{k|k-1} & \triangleq \mathbb{E} \left[ \tilde{x}_{k|k-1}\tilde{x}_{k|k-1}^\top  \right] ,\\
		\nonumber \Sigma _{k|k} & \triangleq \mathbb{E} \left[ \tilde{x}_{k|k}\tilde{x}_{k|k}^\top  \right] \\
		\nonumber   & =\Sigma _{k|k-1}-K_k\left( \breve{C}_k\Sigma _{k|k-1}\breve{C}_k^\top +\breve{R} \right) K_{k}^\top \\
		& ~~~~+K_k\left( \breve{R}_{dec}+s\mathbb{E} \left[ \breve{v}_k\breve{e}_{k}^\top +\breve{e}_k\breve{v}_{k}^\top  \right] \right) K_{k}^\top .
	\end{align}
	
	 The following lemma establishes the boundedness of the term $\breve{R}_{dec}+ s\mathbb{E} \left[ \breve{v}_k\breve{e}_{k}^\top +\breve{e}_k\breve{v}_{k}^\top \right] $.
	
	\begin{lemma}\label{chap04:lemma1}
		Consider the decoding error covariance $\breve{R}_{dec}$ and the covariance term $\mathbb{E} \left[ \breve{v}_k\breve{e}_{k}^\top +\breve{e}_k\breve{v}_{k}^\top  \right] $ introduced by the PPM \eqref{chap04:eq7}-\eqref{chap04:eq9}. The following inequalities hold:
		\begin{align}\label{chap04:eq23}
			& \breve{R}_{dec}+s\mathbb{E} \left[ \breve{v}_k\breve{e}_{k}^\top +\breve{e}_k\breve{v}_{k}^\top  \right] \nonumber \\
			& \le \mathrm{diag} \biggl\{ \gamma _{i,k}^{2}\left( s^2\delta _{N_i}+s\eta _i+\frac{\delta _{N_i}}{s\eta _i} \right) \left( C_i\Sigma _{k|k-1}C_{i}^\top +R_i \right) \biggr\} _{i=1}^{M}  \nonumber \\
			&    \le V \left( \breve{C}_k\Sigma _{k|k-1}\breve{C}_k^\top +\breve{R} \right) V ,
		\end{align}
		where
		\begin{equation}\label{chap04:eq24}
			V\triangleq \mathrm{diag}\left\{ \sqrt{s^2\delta _{N_i}+s\eta _i+\frac{\delta _{N_i}}{s\eta _i}}\cdot I_{d_{y_i}} \right\} _{i=1}^{M}.
		\end{equation}
	\end{lemma}
	
	\begin{IEEEproof}
		For the decoding error of the $i$-th decoder, $e_{dec_{i,k}}$, there exists a distortion rate $\delta _{N_i}\in \left( 0,1 \right) $ such that:
		\begin{align}\label{chap04:eq25}
			& \mathbb{E} \left[ e_{dec_{i,k}}\left( e_{dec_{i,k}} \right) ^\top \right] \nonumber \\
			& =s^2\mathbb{E} \left[ e_{i,k}e_{i,k}^\top  \right] \nonumber \\
			& \le \delta _{N_i}\mathbb{E} \biggl[ \left( y_{i,k}-C_i \hat{x}_{k|k-1} \right) \left( y_{i,k}-C_i\hat{x}_{k|k-1} \right) ^\top \biggr] \nonumber \\
			& =\delta _{N_i}\left( C_i\Sigma _{k|k-1}C_{i}^\top +R_i \right) .
		\end{align}
		As the number of quantization levels $N_i \to \infty$, we have $\delta _{N_i} \to 0$. 
		The decoding error covariance of legitimate users satisfies:
		\begin{align}\label{chap04:eq26}
			\nonumber \breve{R}_{dec} & =\mathrm{diag}\left\{ \left( \gamma _{i,k} \right) ^2s^2\mathbb{E} \left[ e_{i,k}e_{i,k}^\top  \right] \right\} _{i=1}^{M}\\
			&     \le \mathrm{diag}\left\{ \left( \gamma _{i,k} \right) ^2s^2 \delta _{N_i} \left( C_i\Sigma _{k|k-1}C_{i}^\top +R_i \right) \right\} _{i=1}^{M}.
		\end{align}	
		Recalling the basic inequality $xy^\top +yx^\top \le \eta xx^\top +\eta ^{-1}yy^\top $ for a positive scalar $\eta$, we deduce that, for a set of positive scalars $ \left\{\eta_i\right\}_{i=1}^M $, the correlation between the measurement noise and the encoding error satisfies:
		\begin{align}\label{chap04:eq27}
			\mathbb{E} \left[ v_{i,k}e_{i,k}^\top +e_{i,k}v_{i,k}^\top  \right] & \le \eta_i \mathbb{E} \left[ v_{i,k}v_{i,k}^\top  \right] +\frac{1}{\eta_i}\mathbb{E} \left[ e_{i,k}e_{i,k}^\top  \right] \nonumber \\
			&  =\eta _iR_i+\frac{1}{\eta _i}\mathbb{E} \left[ e_{i,k}e_{i,k}^\top  \right] \nonumber \\
			&  \le \eta _i\left( C_i\Sigma _{k|k-1}C_{i}^\top +R_i \right) \nonumber \\
			&  ~~~~+\frac{1}{\eta _i}\frac{1}{s^2}\delta _{N_i} \left( C_i\Sigma _{k|k-1}C_{i}^\top +R_i \right) \nonumber \\
			&   =\left( \eta _i+\frac{\delta _{N_i}}{\eta _is^2} \right) \left( C_i\Sigma _{k|k-1}C_{i}^\top +R_i \right).
		\end{align}
		Subsequently, we derive:
		\begin{align}\label{chap04:eq28}
			&  \mathbb{E} \left[ \breve{v}_k\breve{e}_{k}^\top +\breve{e}_k\breve{v}_{k}^\top  \right]  \nonumber \\
			& =\mathrm{diag}\left\{ \gamma _{i,k}^{2}\mathbb{E} \left[ v_{i,k}e_{i,k}^\top +e_{i,k}v_{i,k}^\top  \right] \right\} _{i=1}^{M}  \nonumber \\
			&  \le \mathrm{diag}\left\{ \gamma _{i,k}^{2}\left( \eta _i+\frac{\delta _{N_i}}{\eta _is^2} \right) \left( C_i\Sigma _{k|k-1}C_{i}^\top +R_i \right) \right\} _{i=1}^{M}.
		\end{align}
		By combining \eqref{chap04:eq26} and \eqref{chap04:eq28}, the proof of the lemma is complete.
	\end{IEEEproof}

	Before proving the boundedness of the estimation error covariance, we introduce several auxiliary definitions. 
	First, the matrix $W_k$ is defined as:
	\begin{align}\label{chap04:eq29}
		& W_kW_k \nonumber \\
		& \triangleq \frac{\lambda _{min}\left\{ \breve{C}_k\Sigma _{k|k-1}\breve{C}_k^\top +\breve{R}-V \left( \breve{C}_k\Sigma _{k|k-1}\breve{C}_k^\top +\breve{R} \right) V  \right\}}{\lambda _{\max}\left\{ \breve{C}_k\Sigma _{k|k-1}\breve{C}_k^\top +\breve{R} \right\}} \nonumber \\
		& ~~~~ \times I_{\Sigma d_y}.
	\end{align}
	Using this definition and Lemma \ref{chap04:lemma1}, we derive that the actual state error covariance is bounded by:
	\begin{align}\label{chap04:eq30}
		\Sigma _{k|k} & =\Sigma _{k|k-1}-K_k \left( \breve{C}_k\Sigma _{k|k-1}\breve{C}_k^\top +\breve{R} \right) K_{k}^\top \nonumber \\
		&  ~~~~ +K_k \left( \breve{R}_{dec}+s\mathbb{E} \left[ \breve{v}_k\breve{e}_{k}^\top +\breve{e}_k\breve{v}_{k}^\top  \right] \right) K_{k}^\top \nonumber \\
		& \le \Sigma _{k|k-1}-K_k \left( \breve{C}_k\Sigma _{k|k-1}\breve{C}_k^\top +\breve{R} \right) K_{k}^\top \nonumber \\
		&  ~~~~ +K_k V  \left( \breve{C}_k\Sigma _{k|k-1}\breve{C}_k^\top +\breve{R} \right) V  K_{k}^\top  \nonumber \\
		& \le \Sigma _{k|k-1}-K_kW_k\left( \breve{C}_k\Sigma _{k|k-1}\breve{C}_k^\top +\breve{R} \right) W_k K_{k}^\top   \nonumber \\
		& \triangleq \bar{\mathscr{U}}_k \ge 0.
	\end{align}
It is therefore sufficient to prove that $ \mathbb{E}\left[ \bar{\mathscr{U}}_k \right] $ is bounded in order to deduce that the expected actual state error covariance, 
 $\mathbb{E}\left[\Sigma _{k|k}\right]$, is also bounded.
	Next, the matrix $W_k$ is decomposed as $W_k=\mathrm{diag}\left\{ W_{i,k}\cdot I_{d_{y_i}} \right\} _{i=1}^{M}$. 
	Define the matrix $\mathscr{S} _{i,k}=\tilde{C}_{i,k}^\top R_{i}^{-1}\tilde{C}_{i,k}$, where $\tilde{C}_{i,k}=C_i W_{i,k}$.
	Using the notation $\mathscr{H} _{i,k}=\sqrt{R_{i}^{-1}}\tilde{C}_{i,k}$, the matrix  $\mathscr{S} _{i,k}$ can be factorized as $\mathscr{S} _{i,k}=\mathscr{H} _{i,k}^\top \mathscr{H} _{i,k}$.
	
	We construct the matrix:
	\begin{align}\label{chap04:eq41}
		\mathcal{W} = & \vec{1}_{\Sigma d_y}\vec{1}_{\Sigma d_y}^\top +\left( \mathrm{diag}\left\{ \frac{\bar{\gamma}_i}{1-\bar{\gamma}_i}\cdot I_{d_{y_i}} \right\} _{i=1}^{M} \right) \nonumber \\
		& \times \left( \mathrm{diag} \left\{ \vec{1}_{d_{y_i}}\vec{1}_{d_{y_i}}^\top  \right\} _{i=1}^{M} \right),
	\end{align}
    where $\vec{1}_{d_{y_i}} \in \mathbb{R}^{d_{y_i}}$ is a column vector with all entries equal to 1, and $\Sigma d_y$ is shorthand for ${\sum_{i=1}^{M} d_{y_i}}$.
	The modified algebraic Riccati equation (MARE) is then defined as:
	\begin{align}\label{chap04:eq31}
		g_{\left\{ \bar{\gamma}_i, \delta_{N_i} \right\} _{i=1}^{M}}\left( X \right) & =AXA^\top +Q-AX \mathscr{H} _{k}^\top \bigg[ \mathcal{W}  \odot \big( \mathscr{H} _kX\mathscr{H} _{k}^\top  \nonumber \\	
		& ~~~~ +I_{\Sigma d_y} \big) \bigg] ^{-1}\mathscr{H} _kXA^\top,
	\end{align}
	where $\mathscr{H} _k\triangleq \left[ \begin{matrix}	\mathscr{H} _{1,k}^\top &		\mathscr{H} _{2,k}^\top &		\cdots&		\mathscr{H} _{M,k}^\top \\\end{matrix} \right] ^\top $, and $\odot $ denotes the Hadamard product.

	The following theorem demonstrates the boundedness conditions and provide an explicit upper bound for the actual error covariance (\ref{chap04:eq22}).
	
	\begin{theorem}\label{chap04:theorem2}
		If $\mathscr{U} _{1}^{-}=\Sigma _{1|0}$, and a sequence $\left\{ \mathcal{V} _k \right\} $ satisfies  $\mathcal{V} _1\ge \mathbb{E} \left[ \mathscr{U} _{1}^{-} \right] $ and $g_{\left\{ \bar{\gamma}_i, \delta_{N_i} \right\} _{i=1}^{M}}\left( \mathcal{V} _k \right) =\mathcal{V} _{k+1}$, then for all $k \ge 0$,
		\begin{equation}\label{chap04:eq32}
			\mathbb{E} \left[ \Sigma _{k|k-1} \right] \le \mathcal{V} _k\,\,for\,\,all\,\,k\ge 0.
		\end{equation}
	\end{theorem}

	\begin{IEEEproof}
		Based on the relationship between the actual error matrix $\Sigma _{k|k}$ and it upper bound $\bar{\mathscr{U}}_{k}$ 
		as shown in the inequality \eqref{chap04:eq30}, we define the following matrices:
		\begin{align}\label{chap04:eq33}
			\mathscr{U} _{k}^{-}\triangleq A \bar{\mathscr{U}}_{k-1} A^\top +Q\ge A\Sigma _{k-1|k-1}A^\top +Q=\Sigma _{k|k-1},
		\end{align}
	    and
		\begin{align}\label{chap04:eq34}
			\mathscr{U} _{k}^{+} \triangleq \mathscr{U} _{k}^{-} - \mathscr{U} _{k}^{-} \breve{C}_k^\top W_k \left( \breve{C}_k \mathscr{U} _{k}^{-} \breve{C}_k^\top +\breve{R} \right) ^{-1} W_k \breve{C}_k \mathscr{U} _{k}^{-}.
		\end{align}
	    Since $\mathscr{U} _{k}^{+} = \left[ \left( \mathscr{U} _{k}^{-} \right) ^{-1} + \sum_{i=1}^M{ \gamma _{i,k}  W_{i,k}C_{i}^\top R_{i}^{-1}C_iW_{i,k}}  \right] ^{-1} $,
		and recalling the definitions of the matrices $W_{i,k}$, $\tilde{C}_{i,k}$ and $\mathscr{S} _{i,k}$, we obtain:
		\begin{align}\label{chap04:eq35}
			\left( \mathscr{U} _{k}^{+} \right) ^{-1} & =\left( \mathscr{U} _{k}^{-} \right) ^{-1}+\sum_{i=1}^M{ \gamma _{i,k}   \tilde{C}_{i,k}^\top R_{i}^{-1}\tilde{C}_{i,k} } \nonumber \\
			& = \left( \mathscr{U} _{k}^{-} \right) ^{-1}+\sum_{i=1}^M{ \gamma _{i,k}   \mathscr{S} _{i,k} }.
		\end{align}
		The second term in (\ref{chap04:eq35}) can be rewritten as $\sum_{i=1}^M{ \gamma _{i,k}   \mathscr{S} _{i,k} } = \mathscr{H} _{k}^\top \Gamma _k\mathscr{H} _k $.
		Consequently, $\mathscr{U} _{k+1}^{-} =A\mathscr{U} _{k}^{+}A^\top +Q$ can be expanded as:
		\begin{align}\label{chap04:eq36}
			\mathscr{U} _{k+1}^{-}
			& = A\left[ \left( \mathscr{U} _{k}^{-} \right) ^{-1}+\mathscr{H} _{k}^\top \Gamma _k\mathscr{H} _k \right] ^{-1} A^\top +Q \nonumber \\
			& = A\left[ \left( \mathscr{U} _{k}^{-} \right) ^{-1}+\mathscr{H} _{k}^\top \Gamma _kI_{\Sigma d_y}\Gamma _k\mathscr{H} _k \right] ^{-1}A^\top +Q \nonumber \\
			& = A\mathscr{U} _{k}^{-}A^\top +Q-A\mathscr{U} _{k}^{-}\mathscr{H} _{k}^\top \Gamma _k \big( \Gamma _k\mathscr{H} _k\mathscr{U} _{k}^{-}\mathscr{H} _{k}^\top \Gamma _k \nonumber  \\
			& ~~~~ +I_{\Sigma d_y} \big) ^{-1}\Gamma _k\mathscr{H} _k\mathscr{U} _{k}^{-}A^\top \nonumber  \\
			&   = A\mathscr{U} _{k}^{-}A^\top +Q-A\mathscr{U} _{k}^{-}\mathscr{H} _{k}^\top \Gamma _k \big( \Gamma _k\mathscr{H} _k\mathscr{U} _{k}^{-} \mathscr{H} _{k}^\top \Gamma _k  \nonumber  \\
			& ~~~~ + \Gamma _k \big) ^{\dagger}\Gamma _k\mathscr{H} _k\mathscr{U} _{k}^{-}A^\top,
		\end{align}
	   where $\dagger$ represesnts the Moore-Penrose pseudoinverse. 
	   The third equality follows from the Woodbury matrix identity, while the last equality is derived from the symmetry and idempotency of the matrix $\Gamma _k$.
		
		By defining the operator $\ell _k\left( K,X,\Gamma \right) =  \left( A-K\Gamma \mathscr{H} _k \right) X \left( A-K\Gamma \mathscr{H} _k \right) ^\top +Q+K\Gamma I\Gamma K^\top$, we prove the theorem using mathematical induction.
		For $k=0$, $\mathbb{E} \left[ \Sigma _{1|0} \right] =\mathbb{E} \left[ \mathscr{U} _{1}^{-} \right] \le \mathcal{V} _1$ holds.
		Assume that for $k=k^{\star}$, $\mathbb{E} \left[ \Sigma _{k^{\star} | k^{\star}-1} \right] \le \mathcal{V} _{k^{\star}} $.
		For the time slot $k= k^{\star} +1$, taking expectation on the right-hand side of (\ref{chap04:eq36}) with respect to $\Gamma _{k^{\star}} $, we have
		\begin{align}\label{chap04:eq42}
			\mathbb{E} \left[ \mathscr{U} _{k^{\star} +1}^{-} \right]  
			& \le \mathbb{E} \left[ \ell _{k^{\star} }\left( \bar{\mathcal{K}}_{k^{\star} },\mathscr{U} _{k^{\star} }^{-},\Gamma_{k^{\star} } \right) \right] \nonumber \\
			& =\mathbb{E} \biggl\{ A \mathscr{U} _{k^{\star}}^{-} A^\top +Q - \bar{\mathcal{K}}_{k^{\star}} \bar{\Gamma} \big[ \mathcal{W} _{k^{\star}} \odot \big( \mathscr{H} _{k^{\star}} \mathscr{U} _{k^{\star}}^{-}  \nonumber  \\
			& ~~~~ \times \mathscr{H} _{k^{\star}}^\top  +I_{\Sigma d_y} \big) \big] \bar{\Gamma} \bar{\mathcal{K}}_{k^{\star}} ^\top - A \mathscr{U} _{k^{\star}}^{-} \mathscr{H} _{k^{\star}} ^\top \bar{\Gamma} \bar{\mathcal{K}}_{k^{\star}} ^\top  \nonumber \\
			& ~~~~ - \bar{\mathcal{K}}_{k^{\star}} \bar{\Gamma} \mathscr{H} _{k^{\star}} \mathscr{U} _{k^{\star}}^{-} A^\top \biggr\} \nonumber \\
			& = \mathbb{E} \left[ g_{\left\{ \bar{\gamma}_i, \delta_{N_i} \right\} _{i=1}^{M}}\left( \mathscr{U} _{k^{\star}}^{-} \right) \right],
		\end{align}
        where
        \begin{align}\label{chap04:eq39}
        	\bar{\mathcal{K}}_{k^{\star}} \triangleq A\mathscr{U} _{k^{\star}}^{-}\mathscr{H} _{k^{\star}}^\top \left[ \mathcal{W} _{k^{\star}} \odot \left( \mathscr{H} _{k^{\star}} \mathscr{U} _{k^{\star}}^{-} \mathscr{H} _{k^{\star}}^\top +I_{\Sigma d_y} \right) \right] ^{-1} \bar{\Gamma}^{-1},
        \end{align}
        and
        \begin{align}\label{chap04:eq40}
        	\bar{\Gamma} \triangleq \mathrm{diag}\left\{ \bar{\gamma}_i\cdot I_{d_{y_i}} \right\} _{i=1}^{M}.
        \end{align}
		Due to the concavity and the increasing monotonicity of $g_{\left\{ \bar{\gamma}_i, \delta_{N_i} \right\} _{i=1}^{M}}\left( X \right)$, 
		we have
		\begin{align}
			\mathbb{E} \left[ g_{\left\{ \bar{\gamma}_i, \delta_{N_i} \right\} _{i=1}^{M}}\left( \mathscr{U} _{k^{\star}}^{-} \right) \right] & \le g_{\left\{ \bar{\gamma}_i, \delta_{N_i} \right\} _{i=1}^{M}}\left( \mathbb{E} \left[ \mathscr{U} _{k^{\star}}^{-} \right] \right) \nonumber \\
			& \le g_{\left\{ \bar{\gamma}_i, \delta_{N_i} \right\} _{i=1}^{M}}\left( \mathcal{V} _{k^{\star}} \right) =\mathcal{V} _{k^{\star} +1}.
		\end{align}
		Therefore, by mathematical induction, $\mathbb{E} \left[ \Sigma _{k|k-1} \right] \le \mathcal{V} _k$ holds for all $ k \ge 0$.
	\end{IEEEproof}
	
	Next, we discuss the conditions for the uniqueness of the solution to $g_{\left\{ \bar{\gamma}_i, \delta_{N_i} \right\} _{i=1}^{M}}\left( \mathcal{V} _k \right) =\mathcal{V} _{k+1}$ and the convergence of $\left\{ \mathcal{V} _k \right\} $. 
	
	\begin{theorem}\label{chap04:theorem3}
		For the sequence $\left\{ \mathcal{V} _k \right\} $, $\mathcal{V} _{k} = g_{\left\{ \bar{\gamma}_i, \delta_{N_i} \right\} _{i=1}^{M}}\left( \mathcal{V} _k \right)$ has a unique solution under arbitrary initial conditions if:
		(1) There exist $\breve{\Sigma}\ge 0$, $\mathcal{K} $ and $\left\{ \bar{\gamma}_i \right\} _{i=1}^{M}$ such that:
		\begin{align}\label{chap04:eq47}
			& \breve{\Sigma} > \left( A-\mathcal{K} \bar{\Gamma} \mathscr{H} _k \right) \breve{\Sigma}\left( A-\mathcal{K} \bar{\Gamma} \mathscr{H} _k \right) ^\top \nonumber \\
			& +\mathcal{K} \left\{ \left[ \breve{R}_{\gamma _k}\left( \mathrm{diag}\left\{ \vec{1}_{d_{y_i}}\vec{1}_{d_{y_i}}^\top  \right\} _{i=1}^{M} \right) \right] \odot \left( \mathscr{H} _k\breve{\Sigma}\mathscr{H} _{k}^\top  \right) \right\} \mathcal{K} ^\top,
		\end{align}
		where $\breve{R}_{\gamma _k}\triangleq \mathrm{diag}\left\{ \bar{\gamma}_i\left( 1-\bar{\gamma}_i \right) I_{d_{y_i}} \right\} _{i=1}^{M}$ and $\bar{\Gamma} \triangleq \mathrm{diag}\left\{ \gamma _{i,k}\cdot I_{d_{y_i}} \right\} _{i=1}^{M}$. 
		
		(2) Let $\breve{Q}\breve{Q}^\top =\mathcal{B}$. For all $\omega \in \mathbb{R} $, $\left[ \begin{matrix}
			A-e^{j\omega}I_{d_x} & \mathcal{B} \end{matrix}\right]$ is full row rank.
	\end{theorem}
	
	\begin{IEEEproof}
		Following the literature \cite{7997940,Zheng2014_O}, the MARE $g_{\left\{ \bar{\gamma}_i, \delta_{N_i} \right\} _{i=1}^{M}}\left( X \right) =X$ has a unique positive semi-definite solution if the following two conditions are satisfied: 
		
		(1) For the stochastic system:
		\begin{equation}\label{chap04:eq50}
			\tilde{x}_{k+1}=\left( A-\mathcal{K} \Gamma _k\mathscr{H} _k \right) ^\top \tilde{x}_k,
		\end{equation}
	 there exists a static gain $\mathcal{K} $ such that $\lim_{k\rightarrow \infty} \mathbb{E} \left[ \tilde{x}_k \tilde{x}_{k}^\top  \right] =0_{d_x} $, where $0_{d_x}$ is a $d_x$-dimentional matrix with all entries equal to zero.
	
	(2) For the system:
	\begin{align}\label{chap04:eq51}
		\begin{cases}
			x_{k+1} & =A^\top x_k,\\
			y_{k+1} & =\mathcal{B}^\top x_k,
		\end{cases} 			
	\end{align}
	there are no unobservable eigenvalues located on the unit circle.
	
	    For the system (\ref{chap04:eq50}), the covariance is given by
	    \begin{align} \label{eq51mod}
	    	& \mathbb{E} \left[ \tilde{x}_{k+1} \tilde{x}_{k+1}^\top \right] 
	    	= \mathbb{E} \left[ \left( A-\mathcal{K} \Gamma _k\mathscr{H} _k \right) ^\top \tilde{x}_k \tilde{x}_k^\top \left( A-\mathcal{K} \Gamma _k\mathscr{H} _k \right)  \right] \nonumber \\
	    	& = A^\top \mathbb{E} \left[ \tilde{x}_k \tilde{x}_k^\top \right]  A
	    	+ \mathscr{H} _k^\top \left[ \bar{\Gamma}  \odot \left( \mathcal{K}^\top \mathbb{E} \left[ \tilde{x}_k \tilde{x}_k^\top \right] \mathcal{K} \right) \right] \mathscr{H} _k \nonumber \\
	    	& ~~~ - A^\top \mathbb{E} \left[ \tilde{x}_k \tilde{x}_k^\top \right] \mathcal{K} \bar{\Gamma} \mathscr{H} _k 
	    	 - \mathscr{H} _k^\top \bar{\Gamma} \mathcal{K}^\top  \mathbb{E} \left[ \tilde{x}_k \tilde{x}_k^\top \right] A 	 \nonumber \\
	    	& =  \left( A-\mathcal{K} \bar{\Gamma} \mathscr{H} _k \right) ^\top\mathbb{E} \left[ \tilde{x}_k \tilde{x}_k^\top \right]  \left( A-\mathcal{K} \bar{\Gamma} \mathscr{H} _k \right) 
	    	 + \mathscr{H} _k^\top \big[ \bar{\Gamma} 	 \nonumber \\
	    	& ~~~   \odot \left( \mathcal{K}^\top \mathbb{E} \left[ \tilde{x}_k \tilde{x}_k^\top \right] \mathcal{K} \right) \big]  \mathscr{H} _k 
	    	- \mathscr{H} _k^\top \bar{\Gamma} \mathcal{K}^\top  \mathbb{E} \left[ \tilde{x}_k \tilde{x}_k^\top \right] \mathcal{K} \bar{\Gamma} \mathscr{H}_k.
	    \end{align}
        Since the last two terms can be rewritten as 
        $ \mathscr{H} _k^\top \left[ \bar{\Gamma}  \odot \left( \mathcal{K}^\top \mathbb{E} \left[ \tilde{x}_k \tilde{x}_k^\top \right] \mathcal{K} \right) \right] \mathscr{H} _k = \sum_{i=1}^{M} \big( \bar{\gamma}_{i} \mathscr{H} _{i,k}^\top \mathcal{K}^\top $ $ \mathbb{E} \left[ \tilde{x}_k \tilde{x}_k^\top \right] \mathscr{H} _{i,k} \big) $ and 
        $\mathscr{H} _k^\top \bar{\Gamma} \mathcal{K}^\top  \mathbb{E} \left[ \tilde{x}_k \tilde{x}_k^\top \right] \mathcal{K} \bar{\Gamma} \mathscr{H}_k  = \sum_{i=1}^{M} \big( \bar{\gamma}_{i}^2 $ $ \mathscr{H} _{i,k}^\top  \mathcal{K}^\top \mathbb{E} \left[ \tilde{x}_k \tilde{x}_k^\top \right] \mathscr{H} _{i,k} \big)$,
        we substitudte them into (\ref{eq51mod}) to derive the right-hand side of (\ref{chap04:eq47}). 
		Similar to the results in \cite{7997940}, $\lim_{k\rightarrow \infty} \mathbb{E} \left[ \tilde{x}_{k+1} \tilde{x}_{k+1}^\top \right] =0_{d_x} $ is equavlent to that there exists $\breve{\Sigma} >0$, $\mathcal{K}$ and $\bar{\Gamma}$ such that (\ref{chap04:eq47}) is satisfied.
		
		For the system (\ref{chap04:eq51}), if there are no unobservable eigenvalues located on the unit circle, then there does not exist any non-zero vector $\vec{\alpha}$ such that $\left[ \begin{matrix} A-e^{j\omega}I_{d_x} & \mathcal{B} \end{matrix} \right]$ for all $\omega$. This indicates the condition (2) in Theorem \ref{chap04:theorem3}.
		
	\end{IEEEproof}
	In the following corollary, we provide a condition on the convergence of $\mathbb{E} \left[ \Sigma _{k|k-1} \right]$ with respect to channel capacities. 
	We definine the Mahler measure and topological entropy of the system parameter matrix $A$ as $ \mathcal{M} \left( A \right)   \triangleq \prod_{i=1}^{d_x}{\max \left\{ \left| \lambda _{i}\left( A \right) \right|,1  \right\}} $ and $ \bar{h}\left( A \right)  =\ln \mathcal{M} \left( A \right) $,
	where $\lambda _{i}\left( A \right) $ is the $i$-th eigenvalue of $A$. 
	
	\begin{corollary} \label{corollary1}
		Suppose that Condition (2) in Theorem \ref{chap04:theorem3} holds.
		If the total channel capacity $\mathscr{C}$ satisfies
		\begin{equation}\label{chap04:eq48}
			\mathscr{C} >\bar{h}\left( A \right) ,
		\end{equation}
		then there exists a set $\left\{ \bar{\gamma}_i \right\} _{i=1}^{M}$ such that, for any initial value, the upper bound of the expected state estimation error covariance, $\sup \mathbb{E} \left[ \Sigma _{k|k-1} \right] =\mathcal{V} _k $, converges.
		Furthermore, there exists $\bar{\mathcal{V}}$ that satisfies $\bar{\mathcal{V}} = g_{\{\bar{\gamma}_i, \delta_{N_i}\}_{i=1}^{M}}(\bar{\mathcal{V}})$ and 
		\begin{equation}\label{chap04:eq49}
			\lim_{k\rightarrow \infty} \mathcal{V} _k \le \bar{\mathcal{V}}.
		\end{equation}		
	\end{corollary}

	\begin{IEEEproof}

		A similar method to that in \cite{7997940} is employed to transform the system matrix for  further convergence analyasis.		
		Assume that $\left( A,\mathscr{H} _k \right) $ is observable and $\mathscr{H} _k$ has row rank $d_h$.
		A subset of rows is chosen from each $\mathscr{H} _{i,k}$, denoted as $\left\{ \mathscr{H} _{i,k} \left( 1 \right) ,\mathscr{H} _{i,k}\left( 2 \right) ,\cdots ,\mathscr{H} _{i,k}\left( \bar{d}_{y_i} \right) \right\} $, where $0\le \bar{d}_{y_i}\le d_{y_i}$. 
		The selected rows are then used to form a new matrix $\bar{\mathscr{H}}_{i,k}=\left[ \begin{matrix}	\left( \mathscr{H} _{i,k}\left( 1 \right) \right) ^\top &		\left( \mathscr{H} _{i,k}\left( 2 \right) \right) ^\top &		\cdots&		\left( \mathscr{H} _{i,k}\left( \bar{d}_{y_i} \right) \right) ^\top \\\end{matrix} \right] ^\top $, which is combined to construct a full row-rank matrix
		$\bar{\mathscr{H}}_k=\left[ \begin{matrix}	 \bar{\mathscr{H}}_{1,k}^\top &		 \bar{\mathscr{H}}_{2,k}^\top &		\cdots&		 \bar{\mathscr{H}}_{M,k}^\top \\\end{matrix} \right] ^\top $ with observable $\left( A,\bar{\mathscr{H}}_k \right) $ .
		Based on the Wonham decomposition \cite{Wonham1967_O}, there exists a similarity transformation matrix that transforms the matrices $A$ and $\bar{\mathscr{H}}_k$ into the following forms:
		\begin{align}\label{chap04:eq52}
			& \breve{A}=\left[ \begin{matrix}	\breve{A}_1&		0&		\cdots&		0\\	\star&		\breve{A}_2&		\cdots&		0\\	\vdots&		\vdots&		\ddots&		0\\	\star&		\star&		\cdots&		\breve{A}_M\\\end{matrix} \right] , \nonumber \\
			& \breve{\mathscr{H}}_k=\left[ \begin{matrix}	\breve{\mathscr{H}}_{1,k}&		0&		\cdots&		0\\	\star&		\breve{\mathscr{H}}_{2,k}&		\cdots&		0\\	\vdots&		\vdots&		\ddots&		0\\	\star&		\star&		\cdots&		\breve{\mathscr{H}}_{M,k}\\\end{matrix} \right],
		\end{align}
		where
		\begin{align}\label{chap04:eq53}
			& \breve{A}_{i}=\left[ \begin{matrix}	\breve{A}_{i}\left( 1 \right)&		0&		\cdots&		0\\	\star&		\breve{A}_{i}\left( 2 \right)&		\cdots&		0\\	\vdots&		\vdots&		\ddots&		0\\	\star&		\star&		\cdots&		\breve{A}_{i}\left( \bar{d}_{y_i} \right)\\\end{matrix} \right] ,  \nonumber \\
			& \breve{\mathscr{H}}_{i,k}=\left[ \begin{matrix}	\breve{\mathscr{H}}_{i,k}\left( 1 \right)&		0&		\cdots&		0\\	\star&		\breve{\mathscr{H}}_{i,k}\left( 2 \right)&		\cdots&		0\\	\vdots&		\vdots&		\ddots&		0\\	\star&		\star&		\cdots&		\breve{\mathscr{H}}_{i,k}\left( \bar{d}_{y_i} \right)\\\end{matrix} \right] ,
		\end{align}
		in which for $\sum_{i=1}^M{\sum_{\bar{j}=1}^{\bar{d}_{y_i}}{d_{i\bar{j},x}}}=d_x$, 
		$\breve{A}_{i}\left( \bar{j} \right) \in \mathbb{R} ^{d_{i\bar{j},x}\times d_{i\bar{j},x}}$,
		$\breve{\mathscr{H}}_{i,k}\left( \bar{j} \right) \in \mathbb{R} ^{d_{i\bar{j},x}\times 1}$, and each $\left( \breve{A}_{i}\left( \bar{j} \right) ,\breve{\mathscr{H}}_{i,k}\left( \bar{j} \right) \right) $ being observable.  
		
		Since the total channel capacity $\mathscr{C}$ satisfies $\mathscr{C} > \bar{h}\left( A \right)$, 
		there exists a collection of communication rates $\left\{ \bar{\gamma}_{i} \right\}_{i=1}^M$ such that, for the $i$-th channel, $\mathscr{C}_i > \max_{\bar{j}} \left\{ \bar{h}\left( \breve{A}_{i}\left( \bar{j} \right) \right) \right\}$. 
		According to the proof of [\cite{7997940}, Theorem 5], for all $i$ and $\bar{j}$, there exist $\breve{\mathscr{U}}_{i}\left( \bar{j} \right) > 0$ and $\breve{\mathcal{K}}_{i}\left( \bar{j} \right)$ such that 
		\begin{align} \label{eq54mod}
			\breve{\mathscr{U}}_{i} \left( \bar{j} \right) > &
			\left[ \breve{A}_{i}\left( \bar{j} \right) - \breve{\mathcal{K}}_{i} \left( \bar{j} \right) \bar{\gamma}_i \breve{\mathscr{H}}_{i,k}\left( \bar{j} \right) \right]
			\breve{\mathscr{U}}_{i} \left( \bar{j} \right)
			\bigg[ \breve{A}_{i}\left( \bar{j} \right) - \breve{\mathcal{K}}_{i} \left( \bar{j} \right)   \nonumber \\			
			& \times \bar{\gamma}_i \breve{\mathscr{H}}_{i,k}\left( \bar{j} \right) \bigg] ^\top
			+ \bar{\gamma}_i \left( 1-\bar{\gamma}_i \right) \breve{\mathcal{K}}_{i} \left( \bar{j} \right) 
			\breve{\mathcal{H}}_{i} \left( \bar{j} \right) 
			\breve{\mathscr{U}}_{i} \left( \bar{j} \right) \nonumber \\
			& \times \breve{\mathcal{H}}_{i} \left( \bar{j} \right) ^\top
			\breve{\mathcal{K}}_{i} \left( \bar{j} \right) ^\top
		\end{align}      
	     We define the surplus of \eqref{eq54mod} as follows:
	    \begin{align}\label{eq57mod1}
	    	\breve{\Delta}_{i} \left( \bar{j} \right) = & \breve{\mathscr{U}}_{i} \left( \bar{j} \right)
	    	- 	\left[ \breve{A}_{i}\left( \bar{j} \right) - \breve{\mathcal{K}}_{i} \left( \bar{j} \right) \bar{\gamma}_i \breve{\mathscr{H}}_{i,k}\left( \bar{j} \right) \right]
	    	\breve{\mathscr{U}}_{i} \left( \bar{j} \right) \bigg[ \breve{A}_{i}\left( \bar{j} \right) \nonumber \\
	    	& - \breve{\mathcal{K}}_{i} \left( \bar{j} \right) \bar{\gamma}_i  \breve{\mathscr{H}}_{i,k}\left( \bar{j} \right) \bigg] ^\top
	    	- \bar{\gamma}_i \left( 1-\bar{\gamma}_i \right) \breve{\mathcal{K}}_{i} \left( \bar{j} \right) 
	    	\breve{\mathcal{H}}_{i} \left( \bar{j} \right)  \nonumber \\
	    	& \times \breve{\mathscr{U}}_{i} \left( \bar{j} \right) \breve{\mathcal{H}}_{i} \left( \bar{j} \right) ^\top
	    	\breve{\mathcal{K}}_{i} \left( \bar{j} \right) ^\top
	    \end{align}
	    Since $\breve{\Delta}_{i}\left( \bar{j} \right) > 0$, there exists a constant $\breve{\alpha} > 0$ such that $\breve{\Delta}_{i}\left( \bar{j} \right) > \breve{\alpha} I_{d_x}$.
	    
	    We then prove that, for all $i$, there exist $\breve{\mathscr{U}}_{i} > 0$ and $\breve{\mathcal{K}}_{i}$ such that 
        \begin{align} \label{eq55mod}
        	\breve{\mathscr{U}}_{i} > &
        	\left[ \breve{A}_{i} - \breve{\mathcal{K}}_{i}  \bar{\gamma}_i \breve{\mathscr{H}}_{i,k} \right]
        	\breve{\mathscr{U}}_{i} 
        	\bigg[ \breve{A}_{i} - \breve{\mathcal{K}}_{i}   
        	\bar{\gamma}_i \breve{\mathscr{H}}_{i,k} \bigg] ^\top \nonumber \\
        	& + \bar{\gamma}_i \left( 1-\bar{\gamma}_i \right) \breve{\mathcal{K}}_{i} 
        	\breve{\mathcal{H}}_{i} 
        	\breve{\mathscr{U}}_{i}  
        	\breve{\mathcal{H}}_{i}  ^\top
        	\breve{\mathcal{K}}_{i}  ^\top
        \end{align}
        To prove that inequality \eqref{eq55mod} holds, we apply mathematical induction to show that, for $\bar{j} = 1, \cdots, \bar{d}_{y_i}$,
        \begin{align} \label{eq56mod}
        	\breve{\mathscr{U}}_{i}^{\bar{j}} > &
        	\left[ \breve{A}_{i}^{\bar{j}} - \breve{\mathcal{K}}_{i}^{\bar{j}}  \bar{\gamma}_i \breve{\mathscr{H}}_{i,k}^{\bar{j}} \right]
        	\breve{\mathscr{U}}_{i}^{\bar{j}} 
        	\bigg[ \breve{A}_{i}^{\bar{j}} - \breve{\mathcal{K}}_{i}^{\bar{j}}   
        	\bar{\gamma}_i \breve{\mathscr{H}}_{i,k}^{\bar{j}} \bigg] ^\top \nonumber \\
        	& + \bar{\gamma}_i \left( 1-\bar{\gamma}_i \right) \breve{\mathcal{K}}_{i}^{\bar{j}} 
        	\breve{\mathcal{H}}_{i}^{\bar{j}} 
        	\breve{\mathscr{U}}_{i}^{\bar{j}}  
        	\left(\breve{\mathcal{H}}_{i}^{\bar{j}}\right)  ^\top
        	\left(\breve{\mathcal{K}}_{i}^{\bar{j}}\right)  ^\top
        \end{align}
        where
        	\begin{align}\label{eq57mod}
        	& \breve{A}_{i}^{\bar{j}}=\left[ \begin{matrix}	\breve{A}_{i}\left( 1 \right)& ~&	0\\	&	\ddots &	\\	
        	\star&  ~&	\breve{A}_{i}\left( \bar{j} \right) \end{matrix} \right] ,  \nonumber \\
        	& \breve{\mathscr{H}}_{i,k}^{\bar{j}} =\left[ \begin{matrix}	\breve{\mathscr{H}}_{i,k}\left( 1 \right)& ~&	0\\	& \ddots&\\
        	\star&	~&  \breve{\mathscr{H}}_{i,k}\left( \bar{j} \right) \end{matrix} \right], \nonumber \\
        & \breve{\mathscr{U}}_{i}^{\bar{j}}=\left[ \begin{matrix}	\mathscr{U}_{i}\left( 1 \right)& ~&	0\\	&	\ddots &	\\	
        	0&  ~&	 \mathscr{U}_{i}\left( \bar{j} \right)\end{matrix} \right] ,  \nonumber \\
        & \breve{\mathcal{K}}_{i}^{\bar{j}} =\left[ \begin{matrix}	\breve{\mathcal{K}}_{i} \left( 1 \right)& ~&	0\\	& \ddots&\\
        	0&	~&  \breve{\mathcal{K}}_{i} \left( \bar{j} \right)\end{matrix} \right]
        \end{align}
    
     For $\bar{j} = 1$, we have $\breve{\mathscr{U}}_{i}^1 = \breve{\mathscr{U}}_{i}\left(1\right)$ and $\breve{\mathcal{K}}_{i}^1 = \breve{\mathcal{K}}_{i}\left(1\right)$. Thus, inequality \eqref{eq56mod} holds. Now, suppose that for $\bar{j} = \bar{j}^\star - 1$, inequality \eqref{eq56mod} holds.

        For $\bar{j} = \bar{j}^\star$, we define:
        \begin{align} \label{eq58mod}
        	&\breve{\mathscr{U}}_{i}^{\bar{j}^\star}=\left[ \begin{matrix}	\breve{\mathscr{U}}_{i}^{(\bar{j}^\star -1 )} & 	0\\	
        		0&  \breve{\beta} \breve{\mathscr{U}}_{i}\left( \bar{j} ^\star \right) \end{matrix} \right] , 
        	\breve{\mathcal{K}}_{i}^{\bar{j}^\star}=\left[ \begin{matrix}	\breve{\mathcal{K}}_{i}^{(\bar{j}^\star -1 )} & 	0\\	
        		0&   \breve{\mathcal{K}}_{i}\left( \bar{j} ^\star \right) \end{matrix} \right] , \nonumber \\
        	& \breve{A}_{i}^{\bar{j}^\star}=\left[ \begin{matrix}	\breve{A}_{i}^{(\bar{j}^\star -1 )} & 	0\\	
        		A^\star&   \breve{A}_{i}\left( \bar{j} ^\star \right) \end{matrix} \right], 
        	\breve{\Delta}_{i}^{\bar{j}^\star}=\left[ \begin{matrix}	\breve{\Delta}_{i}^{(\bar{j}^\star -1 )} & 	0\\	
        		0&  \breve{\Delta}_{i}\left( \bar{j} ^\star \right) \end{matrix} \right].
        \end{align}
       Substituting \eqref{eq58mod} into the definition of $\breve{\Delta}_{i}^{\bar{j}}$ for $\bar{j} = \bar{j}^\star$, we have:
       \begin{align}\label{eq59mod}
       	\breve{\Delta}_{i}^{\bar{j}}= & \breve{\mathscr{U}}_{i} ^{\bar{j}}
       	- 	\left[ \breve{A}_{i}^{\bar{j}}- \breve{\mathcal{K}}_{i} ^{\bar{j}} \bar{\gamma}_i \breve{\mathscr{H}}_{i,k}^{\bar{j}} \right]
       	\breve{\mathscr{U}}_{i} ^{\bar{j}} \bigg[ \breve{A}_{i}^{\bar{j}}- \breve{\mathcal{K}}_{i} ^{\bar{j}} \bar{\gamma}_i \breve{\mathscr{H}}_{i,k}^{\bar{j}} \bigg] ^\top \nonumber \\
       	& - \bar{\gamma}_i \left( 1-\bar{\gamma}_i \right) \breve{\mathcal{K}}_{i} ^{\bar{j}}
       	\breve{\mathcal{H}}_{i} ^{\bar{j}}
       	\breve{\mathscr{U}}_{i} ^{\bar{j}}  
       	\left( \breve{\mathcal{H}}_{i} ^{\bar{j}} \right) ^\top
       	\left( \breve{\mathcal{K}}_{i} ^{\bar{j}} \right) ^\top
       \end{align}
       Since there exists $\breve{\alpha} > 0$ such that $\breve{\Delta}_{i}\left(\bar{j}^\star\right) > \breve{\alpha} I_{d_x}$, we select a sufficiently large value for the scaling parameter $\breve{\beta}$ so that:
       \begin{align}\label{eq60mod}
       	& \breve{\beta} \breve{\alpha} I_{d_x} > A^\star \breve{\mathscr{U}}_{i}^{(\bar{j}^\star -1 )} \bigg( \breve{A}_{i}^{(\bar{j}^\star -1 )} -  \breve{\mathcal{K}}_{i}^{(\bar{j}^\star -1 )} 
       	\bar{\gamma}_i \breve{\mathcal{H}}_{i,k}^{(\bar{j}^\star -1 )} \bigg) ^\top \breve{\Delta}_{i}^{\bar{j}^\star -1}    \nonumber \\
       	& \times \bigg( \breve{A}_{i}^{\bar{j}^\star-1 } - \breve{\mathcal{K}}_{i}^{(\bar{j}^\star -1 )} \bar{\gamma}_i \breve{\mathcal{H}}_{i,k}^{(\bar{j}^\star -1 )} \bigg) \bigg( \breve{A}_{i}^{(\bar{j}^\star -1 )}
       	- \breve{\mathcal{K}}_{i}^{(\bar{j}^\star -1 )} \bigg) ^\top 
       	 \nonumber \\
       	& - A^\star \breve{\mathscr{U}}_{i}^{(\bar{j}^\star -1 )}\left( A^\star \right)^\top. 
       \end{align}  
       The right-hand side of inequality \eqref{eq60mod} corresponds to the lower right block of $\breve{\Delta}_{i}^{\bar{j}}$ in \eqref{eq59mod}. 
       Thus, there exist $\breve{\mathscr{U}}_{i}^{\bar{j}^\star} > 0$ and $\breve{\mathcal{K}}_{i}^{\bar{j}^\star}$ such that inequality \eqref{eq56mod} is satisfied for $\bar{j} = 1, \cdots, \bar{d}_{y_i}$. For $\bar{j} = \bar{d}_{y_i}$, we have $\breve{\mathscr{U}}_{i} = \breve{\mathscr{U}}_{i}^{\bar{d}_{y_i}}$ and $\breve{\mathcal{K}}_{i} = \breve{\mathcal{K}}_{i}^{\bar{d}_{y_i}}$, thereby ensuring that inequality \eqref{eq55mod} holds.

   In the similar way, let $\breve{\mathscr{U}} = \mathrm{diag} \left\{ \breve{\mathscr{U}}_{1}, \breve{\beta}_1 \breve{\mathscr{U}}_{2}, \cdots, \breve{\beta}_{M}-1 \breve{\mathscr{U}}_{M} \right\}$  and $\breve{\mathcal{K}}_{i} = \mathrm{diag} \left\{ \breve{\mathcal{K}}_{i} \right\}_{i=1} ^{M}$.
   The following inequality holds for sufficiently large values of $\left\{\breve{\beta}_i\right\}_{i=1}^{M-1}$:
   \begin{align}
   	\breve{\mathscr{U}} > 
   	&  \left( \breve{A} -\breve{\mathcal{K}} \breve{\mathscr{U}} \breve{\mathscr{H}} _k \right) \breve{\mathscr{U}} \left( \breve{A} - \breve{\mathcal{K}} \breve{\mathscr{U}} \breve{\mathscr{H}} _k \right) ^\top + \breve{\mathcal{K}} \bigg\{ \bigg[ \breve{R}_{\gamma _k}  \nonumber \\
   	& \times \left( \mathrm{diag}\left\{ \vec{1}_{\bar{d}_{y_i}}\vec{1}_{\bar{d}_{y_i}}^\top  \right\} _{i=1}^{M} \right) \bigg] \odot \left( \breve{\mathscr{H}} _k \breve{\mathscr{U}} \breve{\mathscr{H}} _{k}^\top  \right) \bigg\} \breve{\mathcal{K}} ^\top \nonumber
   \end{align}
By applying the inverse transformation of the Wonham decomposition, it follows that:
\begin{align}
	\mathscr{U} > 
	&  \left( A -\mathcal{K} \mathscr{U} \mathscr{H} _k \right) \mathscr{U} \left( A - \mathcal{K} \mathscr{U} \mathscr{H} _k \right) ^\top + \mathcal{K} \bigg\{ \bigg[ \breve{R}_{\gamma _k}  \nonumber \\
	& \times \left( \mathrm{diag}\left\{ \vec{1}_{\bar{d}_{y_i}}\vec{1}_{\bar{d}_{y_i}}^\top  \right\} _{i=1}^{M} \right) \bigg] \odot \left( \mathscr{H} _k \mathscr{U} \mathscr{H} _{k}^\top  \right) \bigg\} \mathcal{K} ^\top \nonumber
\end{align}

Finally, by Theorem \ref{chap04:theorem3}, the convergence of $\mathcal{V}_k$ is guaranteed.

	\end{IEEEproof}

	\subsection{Confidentiality Analysis: Eavesdropper's Performance}
	
	This subsection evaluates the effectiveness of the proposed PPM by analyzing the state estimation performance of an eavesdropper. 
	First, the eavesdropper's received data, encoded data, measurement noise and parameter matrix are augmented as follows:
	\begin{align}\label{chap04:eq56}
	& \vec{y}_{k}^{e} \triangleq \left[ \begin{matrix} \gamma _{1,k}^{e} y_{1,k}^{\top} &	\gamma _{2,k}^{e} y_{2,k}^{\top} &	\cdots &	\gamma _{M,k}^{e} y_{M,k}^{\top} \end{matrix} \right] ^{\top},  \nonumber \\
	& \breve{y}_{k}^{e}  \triangleq \left[ \begin{matrix}	\gamma _{1,k}^{e} \left(\bar{y}_{1,k}^{e}\right) ^{\top} &	\gamma _{2,k}^{e} \left(\bar{y}_{2,k}^{e}\right) ^{\top} &	\cdots &	\gamma _{M,k}^{e} \left(\bar{y}_{M,k}^{e}\right) ^{\top}  \end{matrix} \right] ^{\top} , \nonumber \\
	& \breve{v}_{k}^{e} \triangleq \left[ \begin{matrix} \gamma _{1,k}^{e}v_{1,k}^{\top} & \gamma _{2,k}^{e}v_{2,k}^{\top} &	\cdots & \gamma _{M,k}^{e}v_{M,k}^{\top}  \end{matrix} \right]^{\top} , \nonumber \\
	& \breve{C}_k^e\triangleq \left[ \begin{matrix}	\gamma _{1,k}^{e} C_1^{\top} &	\gamma _{2,k}^{e} C_2^{\top} &	\cdots &	\gamma _{M,k}^{e} C_M^{\top}  \end{matrix} \right] ^{\top}.
     \end{align}
    The eavesdropper's prediction and estimation errors are defined as
	$\tilde{x}_{k+1|k}^{e}\triangleq x_{k+1}-\hat{x}_{k+1|k}^{e}$ and $\tilde{x}_{k+1|k+1}^{e}\triangleq x_{k+1}-\hat{x}_{k+1|k+1}^{e}$, respectively.
	Based on these definitions, the dynamics of the eavesdropper's errors are expressed as:
	\begin{align}\label{chap04:eq57}
		\begin{cases}	\tilde{x}_{k+1|k}^{e} &  =A\tilde{x}_{k|k}^{e},\\	
			\tilde{x}_{k+1|k+1}^{e}& =\tilde{x}_{k+1|k}^{e}+K_{k+1}^{e}\left( \breve{y}_{k}^{e}-\breve{C}_k^e\hat{x}_{k+1|k}^{e} \right) \\ 
			& =\tilde{x}_{k+1|k}^{e}+K_{k+1}^{e}\left( \breve{C}_k^e\tilde{x}_{k+1|k}^{e}+\breve{v}_{k}^{e}+\bar{e}_{k+1} \right),\\\end{cases}
	\end{align}
	where $\bar{e}_{k+1}$ represents the eavesdropper's decoding error, expressed as $ \bar{e}_k=\left[ \gamma _{1,k}^{e} \bar{e}_{1,k} ^\top,	\gamma _{2,k}^{e} \bar{e}_{2,k}^\top,	\cdots,	\gamma _{M,k}^{e} \bar{e}_{M,k} ^\top, \right] ^\top $.
	Taking expectations on both sides yields the dynamics of the mean estimation errors as:
	\begin{align}\label{chap04:eq58}
		\begin{cases}	\mathbb{E} \left[ \tilde{x}_{k+1|k}^{e} \right] =A\mathbb{E} \left[ \tilde{x}_{k|k}^{e} \right],\\	
			\mathbb{E} \left[ \tilde{x}_{k+1|k+1}^{e} \right] =\mathbb{E} \left[ \tilde{x}_{k+1|k}^{e} \right]  \\
			~~~~~~~~~~~~~~~~~~~ + K_{k+1}^{e}\left( \mathbb{E} \left[ \breve{C}_k^e\tilde{x}_{k+1|k}^{e} \right] +\mathbb{E} \left[ \bar{e}_{k+1} \right] \right).\\\end{cases}
	\end{align}

By demonstrating that the mean decoding error, $\mathbb{E} \left[ \bar{e}_{k+1} \right] $, does not converge, it follows that the mean estimation error, $\mathbb{E} \left[ \tilde{x}_{k+1|k+1}^{e} \right] $, also diverges, i.e., $\lim_{k\rightarrow \infty} \left\| \mathbb{E} \left[ \tilde{x}_{k+1|k+1}^{e} \right] \right\| =\infty $. This result guarantees the preservation of data confidentiality.
	
	The following preliminary lemma is instrumental in establishing the divergence of the mean estimation error $\mathbb{E} \left[ \tilde{x}_{k|k}^{e} \right]$.
	 
	\begin{lemma}\label{chap04:lemma2}
		For the multi-sensor system \eqref{chap04:eq1} and filtering algorithm \eqref{chap04:eq13}-\eqref{chap04:eq16}, there exists a positive scalar $\bar{\kappa}$ such that $\left( K_{k}^{e} \right) ^\top K_{k}^{e}\ge \bar{\kappa}I$ holds.
	\end{lemma}
	\begin{IEEEproof}
		From the expression for the state estimator gain, we have:
		\begin{align}\label{chap04:eq59}
			\left( K_{k}^{e} \right) ^\top K_{k}^{e}=&\left( \breve{C}_kP_{k|k-1}\breve{C}_k^\top +\breve{R} \right) ^{-1}\breve{C}_kP_{k|k-1}^{e}P_{k|k-1}^{e}\breve{C}_k^\top \nonumber \\
			& \times \left( \breve{C}_kP_{k|k-1}^{e}\breve{C}_k^\top +\breve{R} \right) ^{-1}.
		\end{align}
		Since $P_{k|k-1}^{e}\ge Q\ge \lambda _{\min}\left\{ Q \right\} I$ and the innovation covariance is bounded, we can derive the following inequality:
		\begin{align}\label{chap04:eq60}
			\left( K_{k}^{e} \right) ^\top K_{k}^{e}\ge & \left( \lambda _{\min}\left\{ Q \right\} \right) ^2\lambda _{\min}\left\{ \breve{C}_k\breve{C}_k \right\} \nonumber \\
			& \times \left( \lambda _{\max}\left\{ \breve{C}_kP_{k|k-1}^{e}\breve{C}_k^\top +\breve{R} \right\} \right) ^{-2}.
		\end{align}
		Therefore, by defining
		\begin{align}
			\bar{\kappa}=\left( \lambda _{\min}\left\{ Q \right\} \right) ^2\lambda _{\min}\left\{ \breve{C}_k\breve{C}_k \right\} \left( \lambda _{\max}\left\{ \breve{C}_kP_{k|k-1}^{e}\breve{C}_k^\top +\breve{R} \right\} \right) ^{-2}, \nonumber
		\end{align}
		we can conclude that $\left( K_{k}^{e} \right) ^\top K_{k}^{e}\ge \bar{\kappa}I$.
	\end{IEEEproof}
	
	As shown in \eqref{chap04:eq57}, the PPM introduces a decoding error $ \bar{e}_{k} $, which ``pollutes'' the eavesdropper's estimation results.
	We assume that the eavesdropper has limited resources to identify which channel's output contains the encoded data. Consequently, it utilizes all received data, $ \vec{y}_{k}^{e} $, for fusion estimation. 
	In this context, we define the following events for the $i$-th channels, where $i=1,2,\cdots ,M$, and at specific time slots $\bar{k}_i$, with $\bar{k}\ge 0$: 
	\begin{align}\label{chap04:eq62}
		\mathcal{E} _{C_i} & =\left\{ \gamma _{i,\bar{k}_i}=1,\gamma _{i,\bar{k}_i}^{e}=0 \right\} , \nonumber \\ 
		\mathcal{E} _{W_i} & =\left\{ \gamma _{i,k_i}^{e}=1,\forall k_i>\bar{k}_i \right\}.
	\end{align}
    The event $\mathcal{E} _{C_i}$ is referred to as the critical event for the $i$-th channels, which activates the $i$-th local PPM. 
    In the following  lemma, we will demonstrate the PPM is effective even under the ``worst-case'' scenario, $\mathcal{E} _{C_i}\cap \mathcal{E} _{W_i}$, where, for the $i$-th wiretap channel, the eavesdropper fails to capture data only at $\bar{k}_i$ but successfully receives all subsequent transmissions.
	\begin{lemma}\label{chap04:lemma3}
		Consider the multi-sensor system \eqref{chap04:eq1} with encoded measurements transmitted via \eqref{chap04:eq3}-\eqref{chap04:eq6} to the remote estimator \eqref{chap04:eq13}-\eqref{chap04:eq16}. There exists at least one channel where the events $\mathcal{E} _{C_i}=\left\{ \gamma _{i,\bar{k}_i}=1,\gamma _{i,\bar{k}_i}^{e}=0 \right\} $ and $\mathcal{E} _{W_i}=\left\{ \gamma _{i,k_i}^{e}=1,\forall k_i>\bar{k}_i \right\} $ occur at $\bar{k}_i\ge 0$, i.e., $\mathcal{E} _{C_i}\cap \mathcal{E} _{W_i}$, for $i \in \left\{ 1,2,\cdots ,M \right\}$. Given $a_i>1$, in \eqref{chap04:eq7}-\eqref{chap04:eq9}, the eavesdropper's estimation error diverges: $\lim_{k\rightarrow \infty} \left\| \mathbb{E} \left[ \tilde{x}_{k|k}^{e} \right] \right\| =\infty $.
	\end{lemma}

	\begin{IEEEproof}
		
		At a given time step $\bar{k}_i\ge 0$, the event $\mathcal{E} _{C_i}=\left\{ \gamma _{i,\bar{k}_i}=1,\gamma _{i,\bar{k}_i}^{e}=0 \right\} $ occurs. 
		During this event, the eavesdropper fails to obtain the transmitted data, thereby being unable to accurately reconstruct the measurement information. 
		This failure results in the loss of reference measurements required for decoding in subsequent steps. 
		Consequently, starting from $\bar{k}_i+1$, decoding errors appear in the eavesdropper's results.		
		Under the occurrence of event $\mathcal{E} _{W_i}=\left\{ \gamma _{i,k_i}^{e}=1,\forall k_i>\bar{k}_i \right\} $, the eavesdropper's decoding error for $ k> \bar{k}_i $ can be expressed as:
		\begin{align}\label{chap04:eq63}
			\nonumber \bar{e}_{i,k} & =\bar{y}_{i,k}^{e}-y_{i,k}=\bar{y}_{i,k}^{e}-\bar{y}_{i,k}+e_{i,k}s\\
			&  =\left( a_i \right) ^{k-\bar{k}_i-1}\left( \bar{e}_{i,\bar{k}_i+1}-e_{i,\bar{k}_i+1}s \right) +e_{i,k}s,
		\end{align}
		where $\mathbb{E} \left[ \bar{e}_{i,\bar{k}_i+1}-e_{i,\bar{k}_i+1}s \right]$ does not converge to zero. 
		Therefore, when $a_i>1$, it follows that $\lim_{k\rightarrow \infty} \left\| \mathbb{E} \left[ \bar{e}_{i,k} \right] \right\| =\infty $.
		Furthermore, since
		\begin{align}\label{chap04:eq64}
			\left\| \mathbb{E} \left[ \bar{e}_k \right] \right\| & =\sqrt{\sum_{i=1}^M{\left\| \mathbb{E} \left[ \gamma _{i,k}^{e}\bar{e}_{i,k} \right] \right\| ^2}}=\sqrt{\sum_{i=1}^M{\left\| \bar{\gamma}_i^{e}\mathbb{E} \left[ \bar{e}_{i,k} \right] \right\| ^2}} \nonumber \\
			& =\sqrt{\sum_{i=1}^M{\left( \bar{\gamma}_i^{e} \right) ^2\left\| \mathbb{E} \left[ \bar{e}_{i,k} \right] \right\| ^2}},
		\end{align}
		if $\mathcal{E} _{C_i}\cap \mathcal{E} _{W_i}$ occurs at some $\bar{k}_i\ge 0, i\in \left\{ 1,2,\cdots ,M \right\} $, and if any $a_i>1$, then $\lim_{k\rightarrow \infty} \left\| \mathbb{E} \left[ \bar{e}_k \right] \right\| =\infty $.
		We will then complete the proof via contradiction. 
		Assume that when $\mathcal{E} _{C_i}\cap \mathcal{E} _{W_i}$, $i=1,2,\cdots ,M$, occurs at $\bar{k}_i\ge 0$, the sequence $\left\{ \mathbb{E} \left[ \tilde{x}_{k|k}^{e} \right] \right\} _{k>0}$ is ultimately bounded, i.e., there exists a positive scalar $\tilde{\chi}$ such that $\lim_{k\rightarrow \infty} \left\| \mathbb{E} \left[ \tilde{x}_{k|k}^{e} \right] \right\| \le \tilde{\chi}$.
		From the dynamic relationship of the eavesdropper's estimation error:
		\begin{align}\label{chap04:eq65}
			& \left\| K_{k}^{e}\left( \mathbb{E} \left[ \breve{C}_k^e\tilde{x}_{k+1|k}^{e}+\breve{v}_{k}^{e} \right] +\mathbb{E} \left[ \bar{e}_{k+1} \right] \right) \right\| \nonumber \\
			& \le \left\| \mathbb{E} \left[ \tilde{x}_{k+1|k+1}^{e} \right] \right\| +\left\| A\mathbb{E} \left[ \tilde{x}_{k|k}^{e} \right] \right\| \le \left( 1+\left\| A \right\| \right) \tilde{\chi}.
		\end{align}
		By Lemma \ref{chap04:lemma2}, the following inequality holds:
		\begin{equation}\label{chap04:eq66}
			\bar{\kappa}^{\frac{1}{2}}\left\| \left( \mathbb{E} \left[ \breve{C}_k^e\tilde{x}_{k+1|k}^{e}+\breve{v}_{k}^{e} \right] +\mathbb{E} \left[ \bar{e}_{k+1} \right] \right) \right\| \le \left( 1+\left\| A \right\| \right) \tilde{\chi}.
		\end{equation}
		This implies $\lim_{k\rightarrow \infty} \bar{\kappa}^{\frac{1}{2}}\left\| \left( \mathbb{E} \left[ \breve{C}_k^e\tilde{x}_{k+1|k}^{e}+\breve{v}_{k}^{e} \right] +\mathbb{E} \left[ \bar{e}_{k+1} \right] \right) \right\| $ $\le \left( 1+\left\| A \right\| \right) \tilde{\chi}$, which contradicts $\lim_{k\rightarrow \infty} \left\| \mathbb{E} \left[ \bar{e}_{k+1} \right] \right\| =\infty $. Hence, $\lim_{k\rightarrow \infty} \left\| \mathbb{E} \left[ \tilde{x}_{k|k}^{e} \right] \right\| =\infty $ holds.
		It is concluded that for the encoding algorithm \eqref{chap04:eq7}-\eqref{chap04:eq9} with $a_i>1$, if the eavesdropper uses the received data, , $ \vec{y}_{k}^{e} $,, the sequence $\left\{ \mathbb{E} \left[ \tilde{x}_{k|k}^{e} \right] \right\} _{k>0}$ diverges whenever $\mathcal{E} _{C_i}\cap \mathcal{E} _{W_i}$ occurs at any $\bar{k}_i\ge 0$.
	\end{IEEEproof}

	The above analysis evaluates the estimation performance of the eavesdropper under the ``worst-case'' scenario. Building on this, Theorem \ref{chap04:theorem4} extends the analysis for the general case. Before presenting Theorem \ref{chap04:theorem4}, we introduce the following lemma.
	
	\begin{lemma} \label{lemma4} \cite{huang2025recursiveprivacypreservingestimationmarkov}
		For a probability space $\left( \Omega ,\mathcal{F} , \mathcal{P} \right) $, there exists $\mathcal{J} _1\subseteq \mathcal{F} $ being a $\sigma$-algebra $\sigma \left( \mathcal{F} \right) $ and a random variable $Y\in \,\,\Omega $ with $\mathbb{E} \left[ Y \right] \le \infty $. Then $\mathbb{E} \left[ Y|\mathcal{J} _1 \right] $ is a $\mathcal{J}_1$-measurable function satisfying $\int_{J_1} {\mathbb{E} \left[ Y| \mathcal{J} _1 \right] d \mathcal{P} }=\int_{J_1}{Y\,\,d \mathcal{P} }, J_1\in \mathcal{J} _1$. Let us define that $\mathbb{E} \left[ \tilde{Y}_{\mathcal{J} _i} \right] =\mathbb{E} \left\{ Y-\mathbb{E} \left[ Y|\mathcal{J} _i \right] \right\} , i=1,2$. Then if $\mathcal{J} _2\subseteq \mathcal{J} _1\subseteq \mathcal{F} $, we have 
		\begin{equation}
			\left\| \mathbb{E} \left[ \tilde{Y}_{\mathcal{J} _1}|\mathcal{J} _1 \right] \right\| \le \left\| \mathbb{E} \left[ \tilde{Y}_{\mathcal{J} _2}|\mathcal{J} _1 \right] \right\|.
		\end{equation}
	\end{lemma}
	
	\begin{theorem}\label{chap04:theorem4}
		Consider the multi-sensor system \eqref{chap04:eq1}, where data is transmitted to a remote estimator \eqref{chap04:eq13}-\eqref{chap04:eq16} using channels \eqref{chap04:eq3}-\eqref{chap04:eq6}. If
		the PPM \eqref{chap04:eq7}-\eqref{chap04:eq9} satisfies $a_i>1$ for at least one $i$, and
		 $\mathrm{Pr} \left\{ \gamma _{i,\bar{k}_i}=1,\gamma _{i,\bar{k}_i}^{e}=0 \right\} =1$ for some $\bar{k}_i\ge 0$ and $i\in \left\{ 1,2,\cdots ,M \right\}$,
		then the eavesdropper's estimation error satisfies $\lim_{k\rightarrow \infty} \left\| \mathbb{E} \left[ \tilde{x}_{k|k}^{e} \right] \right\| =\infty $.
	\end{theorem}

	\begin{IEEEproof}
		For $ i=1,2,\cdots ,M $, let $h_{i,0:k}=\left( \gamma _{i,0:k}, \gamma _{i,0:k}^{e} \right) =\left( \gamma _{i,0},\cdots,\gamma _{i,k},\gamma _{i,0}^{e},\cdots,\gamma _{i,k}^{e} \right) $ denote the output of the $i$-th channel, where its values are drawn from the set $\left\{ 0,1 \right\} ^{2k+2}$. Similarly, define a deterministic counterpart $l_{i,0:k}=\left( l_{i,0},\cdots,l_{i,k},l_{i,0}^{e},\cdots,l_{i,k}^{e} \right) $ from the same set. The actual channel outputs $\left\{ h_{i,0:k} \right\} _{i=1}^{M}$ can be represented by $\left\{ l_{i,0:k} \right\} _{i=1}^{M}$. The eavesdropper's acquired information is then expressed as $\mathcal{J} _k\left( \left\{ l_{i,0:k} \right\} _{i=1}^{M} \right) =\left\{ \zeta _{i,t}\left( l_{i,t} \right) : l_{i,t}=1,t\le k \right\} $.
		
		Let $\mathcal{J} _k\left( \left\{ \mathfrak{h} _{i,0:k} \right\} _{i=1}^{M} \right) $ denote the eavesdropper's information in the ``worst-case'' scenario, and $\mathcal{J} _k\left( \left\{ \breve{\mathfrak{h}}_{i,0:k} \right\} _{i=1}^{M} \right) $ represent other cases. The $sigma$-algebra relationship satisfies: $\mathcal{J} _k\left( \left\{ \breve{\mathfrak{h}}_{i,0:k} \right\} _{i=1}^{M} \right) \subseteq \mathcal{J} _k\left( \left\{ \mathfrak{h} _{i,0:k} \right\} _{i=1}^{M} \right) $, where $\mathcal{J} _k\left( \breve{\mathfrak{h}} \right) $ is measurable with respect to $\mathcal{J} _k\left( \mathfrak{h} \right) $. By Lemma \ref{lemma4}, the following inequality holds:
		\begin{align}\label{chap04:eq68}
			&\left\| \mathbb{E} \left[ \tilde{x}_{k|k}^{e} \bigg|\mathcal{J} _k\left( \left\{ \breve{\mathfrak{h}}_{i,0:k} \right\} _{i=1}^{M} \right) \right] \right\|  \nonumber \\
			&\le \left\| \mathbb{E} \left[ \tilde{x}_{k|k}^{e} \bigg|\mathcal{J} _k\left( \left\{ \mathfrak{h} _{i,0:k} \right\} _{i=1}^{M} \right) \right] \right\| .
		\end{align}
		From Lemma \ref{chap04:lemma3}, we have $\lim_{k\rightarrow \infty} \left\| \mathbb{E} \left[ \tilde{x}_{k|k}^{e}|\mathcal{J} _k\left( \tilde{h} \right) \right] \right\| =\infty $. Therefore, it follows that $\lim_{k\rightarrow \infty} \left\| \mathbb{E} \left[ \tilde{x}_{k|k}^{e}|\mathcal{J} _k\left( \breve{h} \right) \right] \right\| =\infty $.
	\end{IEEEproof}

	\section{Simulation Examples}

	In this subsection, simulations for the Internet-based three-tank system model are presented to demonstrate the impact of encoding parameters on the estimation performance of legitimate users and eavesdroppers, illustrating the usability of the proposed privacy-preserving and state fusion co-design method.
	
	A PPFE problem is studied for an Internet-based three-tank system shown as in \cite{He2017_F}. The three-tank system is modelled in the following formula:
	\begin{equation}\label{eq46}
		\left\{
		\begin{aligned}
			& x_{k+1}=Ax_k+Bu_k+Dw_k,\\	
			& y_{i,k}=C_i x_k+E_i v_{i,k},
		\end{aligned}
		\right.
	\end{equation}
	where the input is fixed as $u_k=\left[ \begin{matrix}	3.0\times 10^{-5}&		2.0\times 10^{-5}\\\end{matrix} \right] ^\top$, together with the system parameters and covariance of noises are shown as follows: $A=\big[ 	0.9889 ~ 0.0001 ~ 0.0110;  ~ 0.0001 $ $ ~ 0.9774 ~ 0.0119; ~ 0.0110 ~ 0.0119 ~ 0.9770 \big]$, 
	$B=D=\big[ 64.5993 ~ 0.0015; ~ 0.0015 ~ 64.2236; ~ 	0.3604 ~ 0.3910 \big]$, $C_1=\big[1 ~ 0 ~ 0; ~ 0 ~ 0 ~ 1 \big]$, $C_2=\big[1 ~ 0 ~ 0; ~ 0 ~ 1 ~ 0\big]$, $C_3=\big[0 ~ 1 ~ 0; ~ 0 ~ 0 ~ 1\big]$, $E_1 =E_2 =E_3 =I_{2}$,
	$Q=10^{-10}I_{3}$ and $R=10^{-4}I_{3}$.
	The initial conditions are set as $x_0=\left[ \begin{matrix}	0.3 &0.1&		0.2\\\end{matrix} \right] ^\top$ and $P_0=I_{3}$.
	
	The data reception probabilities of the authorized channels are: $\bar{\gamma}_1=0.9$, $\bar{\gamma}_2=0.95$ and $\bar{\gamma}_3=0.85$.
	The interception probabilities of the wiretap channels are: $\bar{\gamma}^e_1=0.9$, $\bar{\gamma}^e_2=0.85$ and $\bar{\gamma}^e_3=0.95$.
	
\begin{figure}[!t]
		\centering
		\includegraphics[width=0.9\linewidth,scale=1]{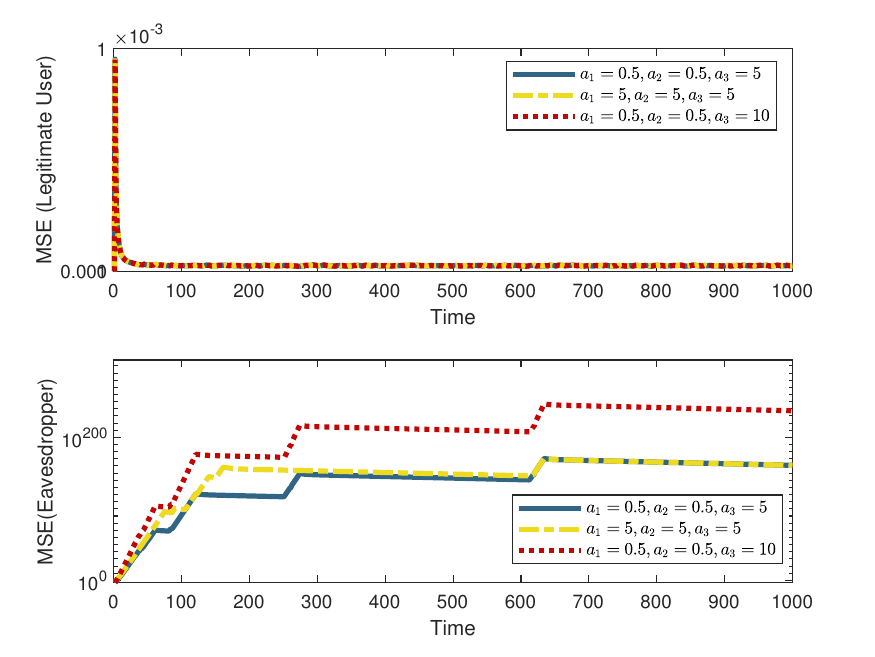}
		\caption{MSE of the legitimate user's and eavesdropper's estimation under different encoding parameters $a_i$.}
		\label{chap04:Fig2}
	\end{figure}

	We first fix the encoding parameters $\delta_i$ and $s$ as $\delta_1 = \delta_2 =\delta_3 =0.01$ and $s=1$, to see the impact of different $a_i$ values on the privacy performance.
	Three parameter groups are tested:
	\begin{equation}\label{chap04:eq:s6}
		\left\{
		\begin{aligned}
			& \text{Group 1: }a_1=0.5,a_2=0.5,a_3=5;\\
			& \text{Group 2: }a_1=0.5,a_2=5,   a_3=5;\\
			& \text{Group 3: }a_1=0.5,a_2=0.5,a_3=10;
		\end{aligned}
		\right.
	\end{equation}
	The comparison results are shown in Fig. \ref{chap04:Fig2}.
	
	From Fig. \ref{chap04:Fig2}, it is observed that the selected parameter groups $a_i$ exhibit no discernible impact on the legitimate user's MSE  but significantly influence the eavesdropper's MSE. 
	The legitimate user's MSE remains unaffected by $a_i$ variations.
	Specifically, the eavesdropper's MSE diverges whenever at least one $a_i > 1$ is present.
	Comparing Groups 1 and 2, increasing the number of active encoding mechanisms (i.e., $a_i > 1$) accelerates the MSE divergence rate but not affect significantly on the magnitude of the eavesdropper's MSE.
	Contrasting Groups 1 and 3, larger $a_i$ values amplify both the divergence rate and magnitude of the eavesdropper's MSE.

	\begin{figure}[!t]
		\centering
		\includegraphics[width=0.9\linewidth,scale=1]{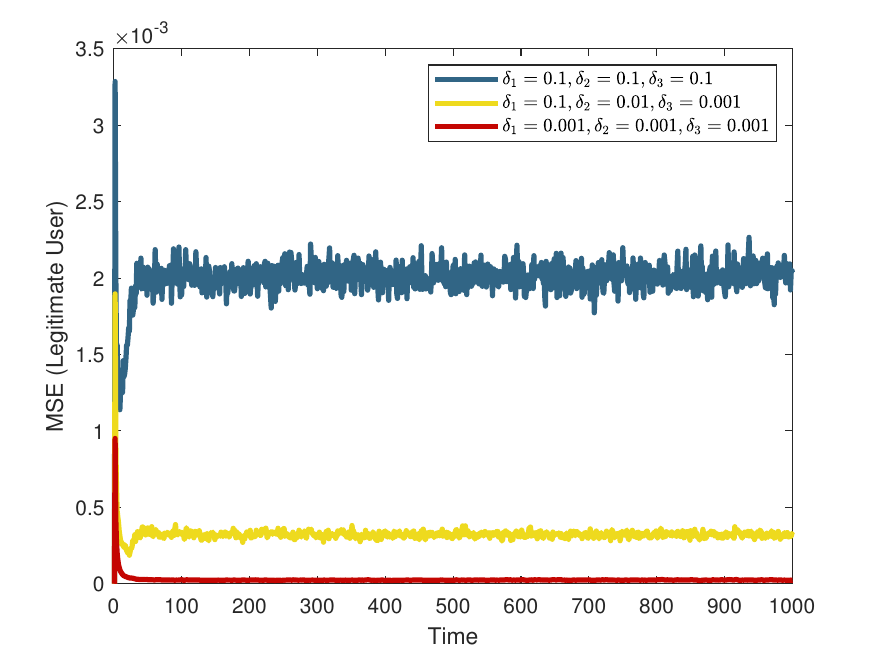}
		\caption{MSE of the legitimate user's estimation under different encoding parameters $\delta_i$.}
		\label{chap04:Fig3}
	\end{figure}

	Then, we fix the encoding parameters $a_i$ and $s$ as $a_1 = a_2 =a_3 =5$ and $s=1$, to see the impact of different $\delta_i$ values on the legitmate user's estimation performance.
	Three $\delta _i$ configurations are tested:
	\begin{equation}\label{chap04:eq:s8}
		\left\{
		\begin{aligned}
			&\text{Group 1: }\delta _1=0.1,    \delta _2=0.1,     \delta _3=0.1;\\
			& \text{Group 2: }\delta _1=0.1,    \delta _2=0.01,   \delta _3=0.001;\\
			& \text{Group 3: }\delta _1=0.001, \delta _2=0.001, \delta _3=0.001;
		\end{aligned}
		\right.
	\end{equation}
	In Fig. \ref{chap04:Fig3}, the MSE of the legitimate user decreases as $\delta _i$ values reduce (red line $<$ yellow line $<$ blue line).
	This confirms that finer encoding precision ($\delta _i \to 0$) enhances estimation accuracy for legitimate users.

	\section{Conclusions}
	This paper has introduced a novel PPFE algorithm for multi-sensor systems under multiple packet dropouts and eavesdropping attacks. 
	By leveraging an encoding-based PPM, the proposed approach has effectively balanced data availability and confidentiality. 
	A modified algebraic Riccati equation, designed to account for the effects of packet dropouts and encoding errors, has ensured the boundedness of the legitimate user's estimation error covariance. 
	Theoretical analysis has showed that the eavesdropper's estimation error diverges when at least one encoding parameter is properly configured ($ a_i > 1 $), thereby providing robust privacy guarantees. 
	Simulation results for an Internet-based three-tank system has further demonstrate the practical applicability of the proposed methodology, highlighting its ability to enhance data privacy without compromising the estimation accuracy of legitimate users.

	\raggedend
	\bibliographystyle{IEEEtran}
	\bibliography{refs}

	\vfill
\end{document}